\documentclass[twocolumn]{aastex63}

          \font\sixrm=cmr6

\newcommand{\sgr}{\mbox{SGR\,J1935+2154~}}
\newcommand{\sgrnos}{\mbox{SGR\,J1935+2154}}

\newcommand{\tbb}{\mbox{$\tau_{\rm bb}$~}}

\def\lambdaC{{\lambda\llap {--}}_{\hbox{\sixrm C}}}

\def\rns{R_{\hbox{\sixrm NS}}}  

\usepackage{amsmath}
\usepackage{amssymb}

\submitjournal{ApJL}

\shorttitle{\sgr bursts}
\shortauthors{Lin et al.}

\graphicspath{{./}{figures/}}

\begin{document}

\title{{\it Fermi}/GBM View of the 2019 and 2020 Burst Active Episodes of \sgrnos}

\correspondingauthor{Lin Lin}
\email{llin@bnu.edu.cn}

\author[0000-0002-0633-5325]{Lin Lin}
\affiliation{Department of Astronomy, Beijing Normal University, Beijing 100875, China}

\author[0000-0002-5274-6790]{Ersin G\"o\u{g}\"u\c{s}}
\affiliation{Sabanc\i~University, Faculty of Engineering and Natural Sciences, \.Istanbul 34956 Turkey}

\author[0000-0002-7150-9061]{Oliver J. Roberts}
\affiliation{Science and Technology Institute, Universities Space and Research Association, 320 Sparkman Drive, Huntsville, AL 35805, USA.}

\author[0000-0003-4433-1365]{Matthew G. Baring}
\affiliation{Department of Physics and Astronomy - MS 108, Rice University, 6100 Main Street, Houston, Texas 77251-1892, USA}

\author[0000-0003-1443-593X]{Chryssa Kouveliotou}
\affiliation{Department of Physics, The George Washington University, 725 21st Street NW, Washington, DC 20052, USA}
\affiliation{Astronomy, Physics, and Statistics Institute of Sciences (APSIS), The George Washington University, Washington, DC 20052, USA}

\author[0000-0002-1861-5703]{Yuki Kaneko}
\affiliation{Sabanc\i~University, Faculty of Engineering and Natural Sciences, \.Istanbul 34956 Turkey}

\author[0000-0001-9149-6707]{Alexander J. van der Horst}
\affiliation{Department of Physics, The George Washington University, 725 21st Street NW, Washington, DC 20052, USA}
\affiliation{Astronomy, Physics, and Statistics Institute of Sciences (APSIS), The George Washington University, Washington, DC 20052, USA}

\author[0000-0002-7991-028X]{George Younes}
\affiliation{Department of Physics, The George Washington University, 725 21st Street NW, Washington, DC 20052, USA}
\affiliation{Astronomy, Physics, and Statistics Institute of Sciences (APSIS), The George Washington University, Washington, DC 20052, USA}

\begin{abstract}

We present temporal and time-integrated spectral analyses of 148 bursts from the latest activation of \sgrnos, observed with {\it Fermi}/GBM from October 4$^{th}$ 2019 through May 20$^{th}$ 2020, excluding a $\sim 130$\,s segment with a very high burst density on April 27$^{th}$ 2020. The 148 bursts presented here, are slightly longer and softer than bursts from earlier activations of \sgrnos, as well as from other magnetars. The long-term spectral evolution trend is interpreted as being associated with an increase in the average plasma loading of the magnetosphere during bursts. We also find a trend of increased burst activity from \sgr since its discovery in 2014. Finally, we find no association of typical radio bursts with X-ray bursts from the source; this contrasts the association of FRB\,200428 with an \sgr X-ray burst, to date unique among the magnetar population.

\end{abstract}


\section{Introduction} \label{sec:intro}

Among the intriguing properties of extremely magnetized neutron stars (a.k.a magnetars,  \cite{dt92,ck1998}), repeated emission of very short, soft $\gamma$-ray bursts is probably their most characteristic attribute (for a review see \cite{kas17}). Burst emission has been detected, at different occurrence rates, from more than two-thirds of the magnetar population \citep{olausen14}. The total energies of these typically short ($\sim0.1$ s) events are very large, ranging anywhere from $\sim$10$^{38}$ erg to 10$^{42}$ erg, and very rarely $\gtrsim$10$^{44}$ erg during the several minute-long Giant Flares (GFs) \citep{hurley99,palmer2005}. 


\sgr was discovered when a short burst triggered the Burst Alert Telescope (BAT) on board the Neil Gehrels Swift Observatory (hereafter {\it Swift}) \citep{mstam14}. Pointed follow-up observations with the {\it Swift}/X-Ray Telescope, Chandra and XMM-Newton
revealed a spin period of 3.24 s and a period derivative of 1.43$\times 10^{-11}$ s/s, therefore, an inferred equatorial surface magnetic field strength of 2.2$\times 10^{14}$ G, thus establishing its magnetar nature \citep{israel16}. Subsequently, \sgr went into multiple short, burst-active episodes in 2015 and 2016, with tens of bursts during each episode \citep{younes17,Lin2020}. From this perspective, \sgr is considered a prolific transient magnetar, according to the classifying scheme of \cite{gogus14}.

In \cite{Lin2020}, we presented a comprehensive investigation of bursts from \sgr during its four active episodes in 2014, 2015 and 2016 (twice), detected with the Gamma-ray Burst Monitor (GBM) on board the {\it Fermi} Gamma-ray Space Telescope ({\it Fermi}) and \textit{Swift}/BAT. During the detailed temporal and spectral analyses of these bursts, we found that the magnetar became more burst-active in every subsequent active episode, emitting 3, 24, 42, and 54 bursts in 2014, 2015, May 2016, and June 2016, respectively. The cumulative energy for each active episode was also observed to grow sequentially over the same time frame; 
$\sim$$1\times10^{39}$, $\sim$$4\times10^{40}$, $\sim$$1\times10^{41}$, $\sim$$4\times10^{41}$ erg, assuming a source distance of 9 kpc. Interestingly, we also found that the spectral behavior of these bursts evolved in time; bursts detected in 2016 were, on average, slightly harder than those in 2014 and 2015. This overall source evolution suggested that the next activation would likely be more intense.

\sgr was active again on October 4$^{th}$ 2019, when it emitted a solitary event. A month later, in November 2019, the source entered a state of heightened activity; this is the first active episode reported in this paper. \sgr returned back to a non-bursting state before resuming activity in late April 2020. There was again, a solitary triggered event in the GBM data on April 10$^{th}$ and one additional event on April 22$^{nd}$ detected with CALET, Konus-Wind and IPN~\citep{GCN27623,GCN27625,GCN27631}; GBM was Earth-occulted during the 
later burst. 
 
On April 27$^{th}$, \sgr entered an extreme burst-active episode emitting hundreds of X-ray bursts over a few minutes \citep{palmer20,younes20}. Strikingly, a bright Fast Radio Burst (FRB~200428) was detected on April 28$^{th}$ from the direction of \sgr \citep{scholz20,bochenek20}, contemporaneous with an X-ray burst from the source \citep{2020ApJ...898L..29M,li2020,kw200428}. \citet{younes20_2} demonstrated that this X-ray burst was spectrally different from all other bursts detected with GBM during the same active episode. 
Following FRB~200428, three weaker radio bursts from \sgr have been reported \citep{fastatel2020b,2020arXiv200705101K}. These were three to six magnitudes dimmer than FRB~200428, each without an X-ray counterpart simultaneously detected
\citep{2020GCN.27679....1L,2020arXiv200705101K}.

In this study, we present detailed temporal and spectral analyses of 148 \sgr bursts detected with GBM during its 2019 (22 bursts) and 2020 (126 bursts) activities, excluding a period with a densely concentrated burst forest, whose analyses 
will be reported elsewhere (Kaneko et al, in preparation). In the following section, we describe our deep search for untriggered bursts from \sgr using the continuous high time resolution data of GBM, and elaborate on our data analysis methodology. We present our results in Section 3, and discuss their implications in Section 4.

\section{Burst Search \& Data Analysis} \label{sec:data}

\sgr is visible for about half of the time by GBM owing to its wide un-occulted field of view, which is afforded by twelve NaI detectors (8\,keV$-$1\,MeV) and two BGO scintillators ($\sim$200\,keV$-\sim$30\,MeV).
A more detailed description of the instrument and scientific data types can be found in \citet{meegan2009}. Our analysis of magnetar bursts, which typically emit at energies $<200$\,keV, is based on the continuous time-tagged event (CTTE) data of NaI detectors, which provides the highest temporal (2 $\mu$s) and spectral (128 channels) resolutions.

We analyzed the data for the 2019 and 2020 outbursts 
in a similar way to our previous studies of the same source \citep{Lin2020}. A Bayesian Block algorithm \citep{Scargle2013} was used to search for magnetar-like short bursts in the CTTE data. 
The algorithm splits up the data into blocks, with each block having a constant rate. This addresses the issue of characterizing any variability in the CTTE data by finding the optimal boundaries between each block, called change points. This allows us to separate statistically significant, valid events, from random noise using a non-parametric light curve analysis \citep{Scargle2013}. The false positive rate of a change point between two blocks was set to 5\% for the entire search, using a prior number of change points through the data~\citep{Scargle2013}. This iterative process is completed when all the parameters from the search are consistent.  
We searched for bursts in the intervals from September 25$^{th}$ 2019 through November 20$^{th}$ 2019 and April 1$^{st}$ 2020 through May 31$^{st}$ 2020. Besides \sgrnos, SGR~1806$-$20 and Swift~J1818.0-1607 were also occasionally active during our search intervals \citep{gcn27672,gcn27696,gcn27746}. All burst candidates found with our Bayesian Block search are localized using the Daughter Of Locburst (DOL) code~\citep{vonKienlin2012}. The average statistical uncertainty at $1\sigma$ confidence level of our sample is $\sim4.7^{\circ}$, and the systematic uncertainty is $\sim4.4^{\circ}$ \citep{Lin2020}. The distance between \sgr and any of the other active magnetars is larger than the location uncertainties. We selected all bursts whose locations on the sky are consistent with \sgrnos. Table \ref{tab:burstlist} lists each burst start time and temporal and spectral characteristics, while Table~\ref{tab:burstepisode} gives a summary of the source activity during each episode.

During the onset of the outburst on April 27$^{th}$, \sgr entered an energetic (fluence $F\sim$2.7$\times 10^{-4}~\rm{erg}~\rm{cm^{-2}}$ in the 8--200\,keV band) period of activity,  lasting $\sim$130~s.
This burst forest was reported by several instruments; it is the first time such behaviour has been observed from \sgr since its discovery. During the forest, the bursts are superimposed on enhanced persistent emission. In this work, we exclude all bursts during this forest (from 18:31:30 to 18:33:40 UTC on 2020 April 27$^{th}$) to keep our sample consistent with that of our previous study \citep{Lin2020}. 
For the bursts in our sample, we ascribe multiple peaks as belonging to the same burst if the time difference between their peaks is less than one quarter of the spin period of \sgrnos, following the convention of \citet{gogus01}. Our final burst sample comprises 148 bursts, of which 22 events were detected late 2019 and 126 early 2020 (see Table \ref{tab:burstlist}). 

\begin{figure*}
\includegraphics*[viewport=50 160 600 495, scale=0.5]{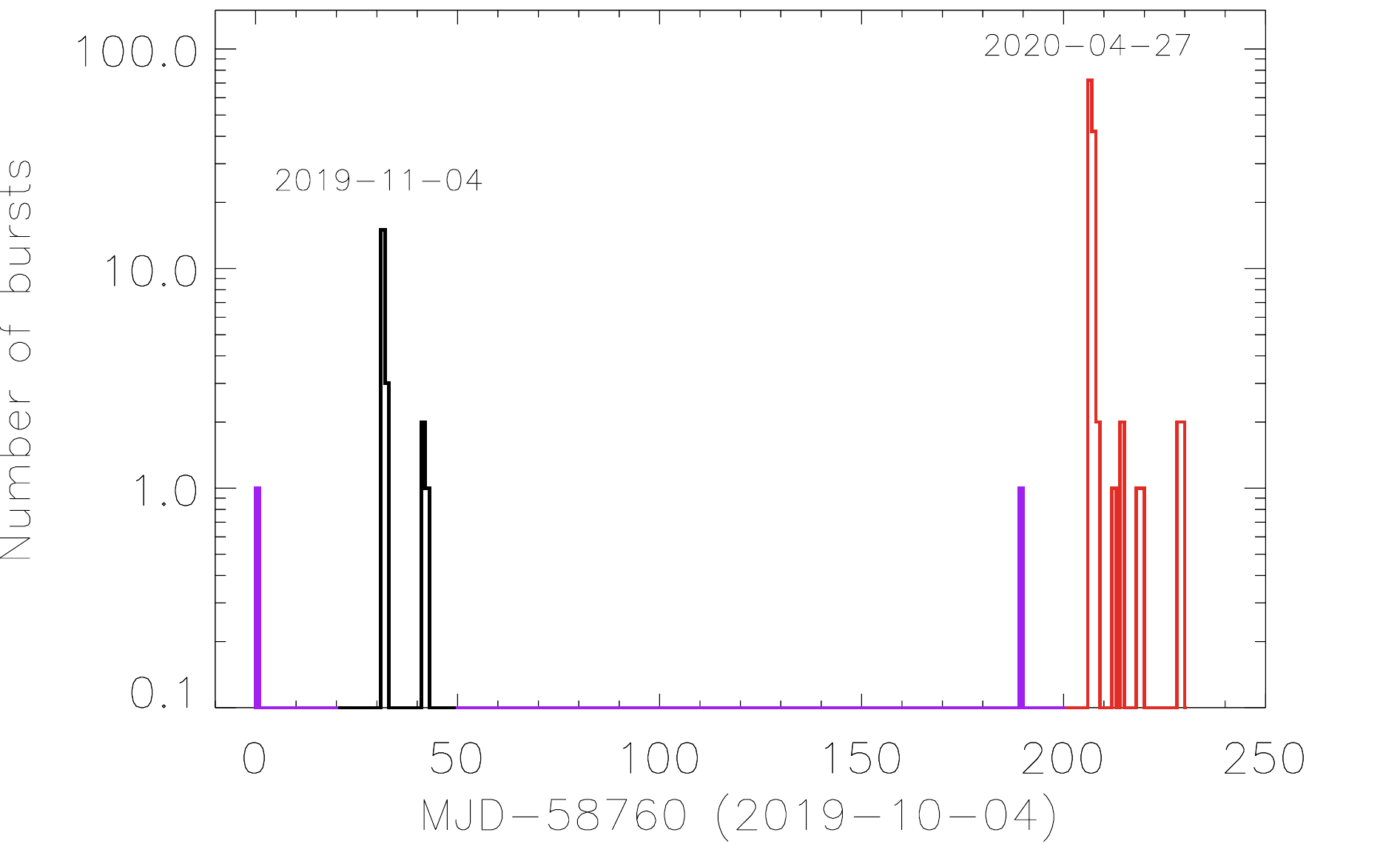}
\includegraphics*[viewport=50 160 600 495, scale=0.5]{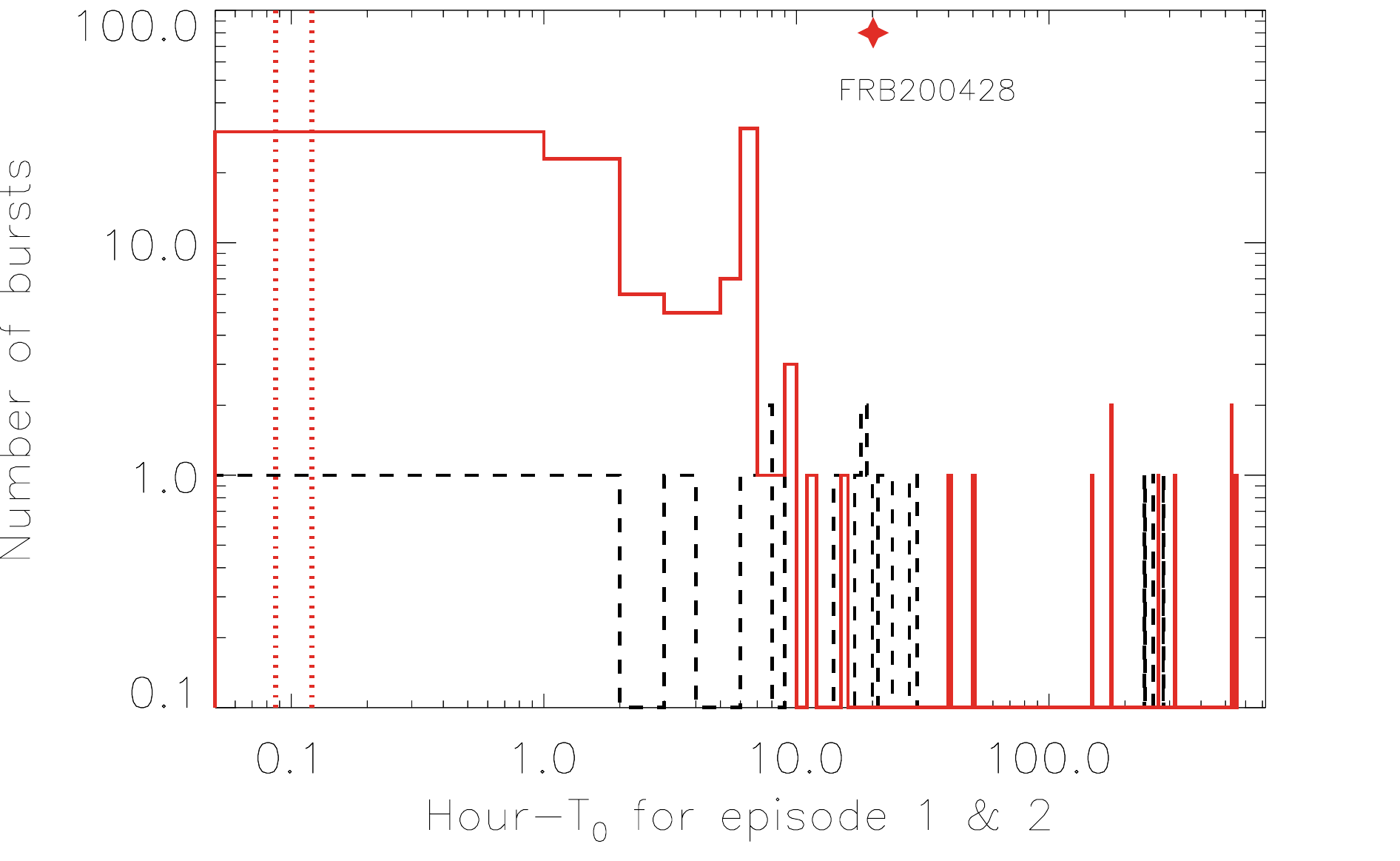}
\caption{\textit{Left:} The burst history of \sgr in 1-day time bins from 2019 October 04 to 2020 May 20. The bursts in episodes 1 and 2 are highlighted in black and red, respectively. Two bursts in purple are isolated events, occurring prior to each episode. 
\textit{Right:} The number of bursts per hour for the first (black dashed line) and second (red solid line) active episode, respectively. The red dotted lines mark the start and stop time of the burst forest not included in this work. The red star shows the relative time of FRB~200428 during the second active episode. We assign black and red in all forthcoming figures to the first and second active episodes, respectively. \label{fig:bursthistory}}
\end{figure*}

 As in \citet{Lin2020}, we define an active bursting episode in this study as a period in which more than two bursts are emitted within 10 days of each other; bursts observed outside this period are excluded. Therefore, we identify two bursting episodes from \sgrnos, which are shown in Figure~\ref{fig:bursthistory}. The properties of these episodes are summarized in Table~\ref{tab:burstepisode}. Note that the two isolated bursts (on October 4$^{th}$ 2019 and April 10$^{th}$ 2020) mentioned in Section \ref{sec:intro} are included in Table \ref{tab:burstepisode} and the whole sample analyses, but are not part of the Episodes 1 and 2 analyses. 
 
\begin{deluxetable*}{lllcccc}
\tablenum{2}
\tablecaption{\sgr Activation Intervals. \label{tab:burstepisode}}
\tablewidth{0pt}
\tablehead{
\colhead{Episode } & \colhead{Start date} & \colhead{End date} & \colhead{Triggered (Untriggered) Events} & \colhead{Total Number} & \colhead{Burst fluence$^{\dagger}$} & \colhead{Burst energy$^{*,\dagger}$} \\
\colhead{} & \colhead{} & \colhead{} & \colhead{}  & \colhead{} & \colhead{($10^{-7}~\rm{erg}~\rm{cm^{-2}}$)} & \colhead{($10^{40}~\rm{erg}$)} 
}
\startdata
1 & 2019 Nov 04 & 2019 Nov 15 & 13(8) & 21 & $127.4\pm0.7$ & $12.3\pm0.1$ \\
2 & 2020 Apr 27 & 2020 May 20$^{\ddag}$ & 28(97) & 125$^{**}$ & $813.3\pm1.7$ & $78.6\pm0.2$ \\ 
all & 2019 Oct 04 & 2020 May 20$^{\ddag}$ & 43(105) & 148$^{**}$ & $968.8\pm1.9$ & $93.6\pm0.2$ \\
\enddata
\tablecomments{$^*$ Assuming a distance of 9 kpc to \sgrnos. \\
$^{**}$ Does not include the bursts from the burst forest.\\
$^{\dagger}$ Values are the sum of fluence and energy in 8$-$200~keV, respectively for all bursts in each episode.\\ 
$^{\ddag}$ The burst search was performed until 2020 May 31. GBM did not trigger on any burst from \sgr after that time. Additional single, untriggered bursts after the end of the 2020 active episodes will not affect our results significantly.}
\end{deluxetable*}

\section{Results} \label{sec:result}

\subsection{Temporal analysis} \label{sec:duration}

\begin{figure*}
\includegraphics*[viewport=75 160 600 495, scale=0.5]{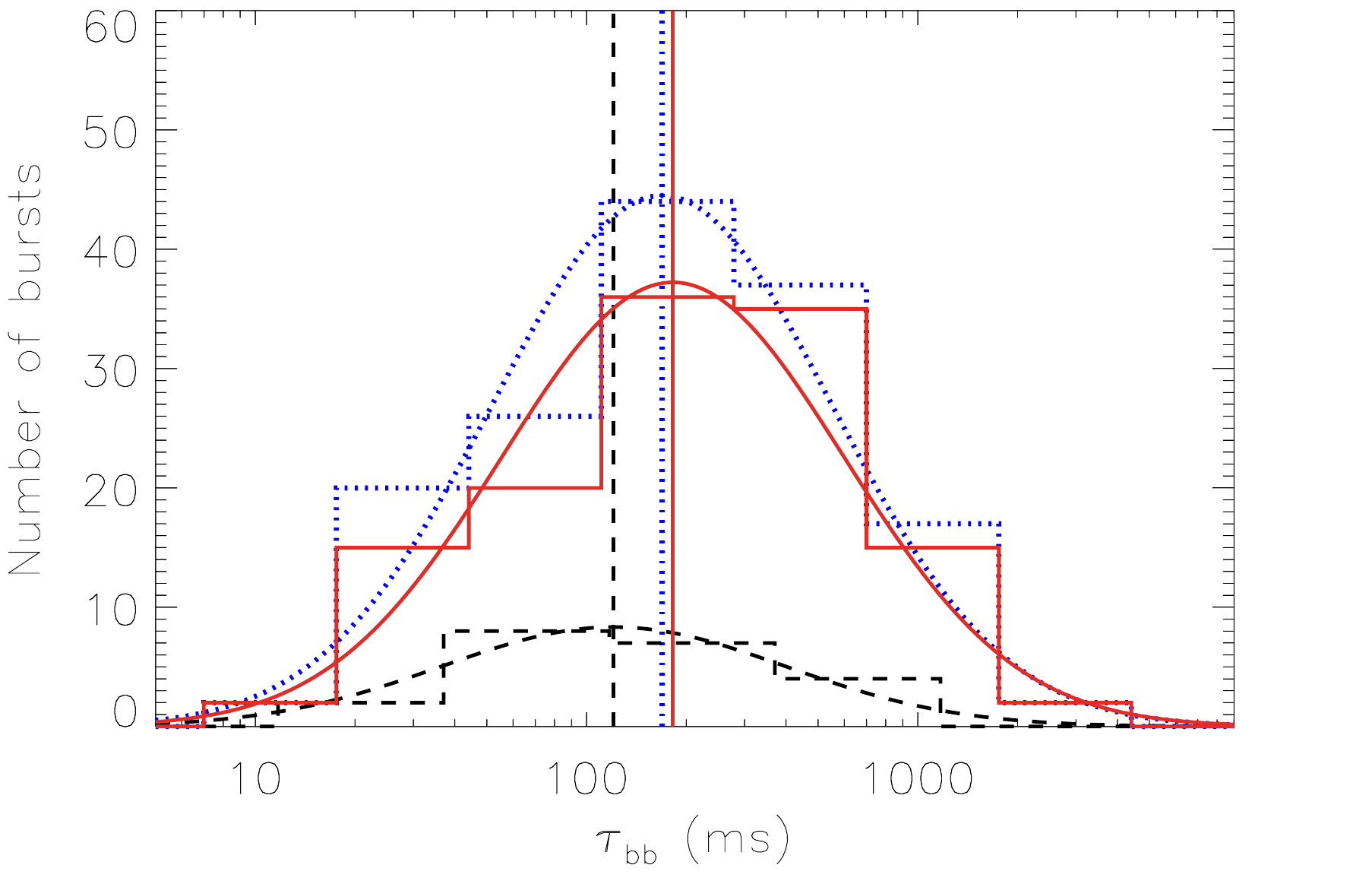}
\includegraphics*[viewport=50 160 600 495, scale=0.5]{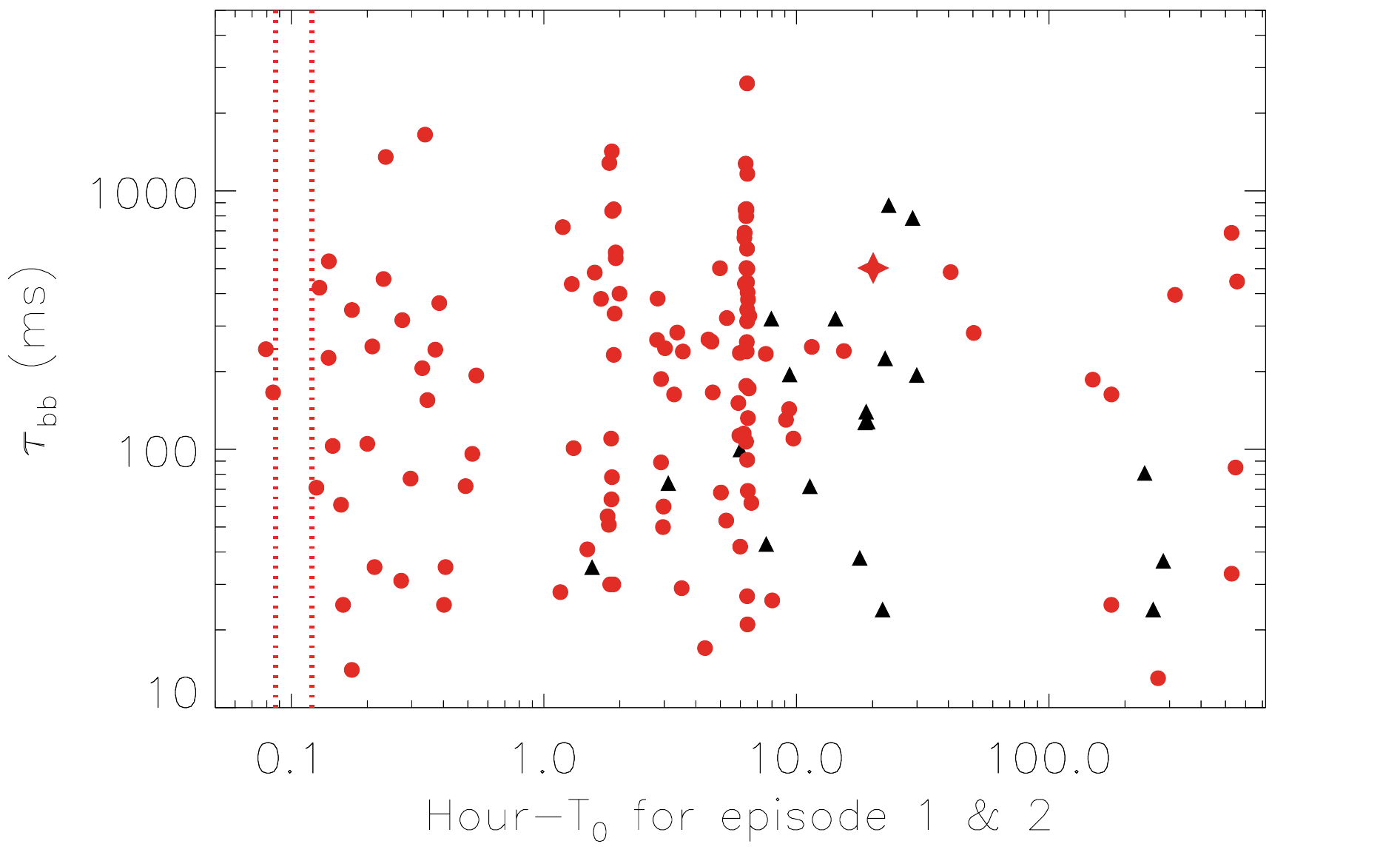}
\caption{\textit{Left:} The distribution of \tbb for the whole sample (blue-dotted histogram), and for active episodes 1 \& 2 (black-dashed and red-solid histograms, respectively). The best fit log-Gaussian functions and corresponding mean values are over-plotted with the same color and style curves and vertical lines, respectively. \textit{Right:} The scatter plot of \tbb 
versus
their start time with respect to the first burst of active episodes 1 (black triangles) and 2 (red dots). The dotted lines mark the start and stop time of the burst forest. The duration and occurrence time of the X-ray burst associated with FRB~200428 is also marked with a red star.
\label{fig:tbb}}
\end{figure*}

The Bayesian block duration (\tbb) is a product of our Bayesian burst search process. It is the total time length of all consecutive Bayesian blocks over the interval of a burst. In this work we calculated \tbb in a similar manner as in \citet{Lin2020}, but with a temporal resolution of 1~ms. We list the \tbb duration of each burst in Table~\ref{tab:burstlist}. 
We find that the distribution of burst durations follows a log-Gaussian trend, as was the case for the duration distributions for \sgr bursts seen prior to 2019, as well as bursts from other magnetars (see e.g., \citealt{Collazzi2015}). We present in the left panel of Figure~\ref{fig:tbb}, the duration distribution along with the best fitting log-Gaussian function, with a mean of $169_{-16}^{+18}$~ms. We also formed separate duration distributions for the 2019 and 2020 episodes and fit them with a log-Gaussian function; we find that the 2020 bursts are slightly longer on average. The cumulative means of the burst durations from 2019 and 2020 are 121 ms and 182 ms, respectively (see Table~\ref{tab:distribution} for details). In the right panel of Figure~\ref{fig:tbb}, we present a scatter plot of \tbb versus burst time, each starting with the first burst of each episode. 
We find that the bursts from the 2020 episode show a significant increase in their frequency of occurrence, during $2-8$ hours after the onset of the episode. Further, in the latter episode, all bursts with \tbb$>1$~s occur within its first ten hours.

\begin{deluxetable*}{lccccccc}
\tablenum{3}
\tablecaption{Results of the Gaussian fits to the temporal and spectral parameter distributions of \sgr bursts. 
\label{tab:distribution}}
\tablewidth{0pt}
\tablehead{
\colhead{Parameter} & \multicolumn{3}{c}{Episode 1} & \colhead{} & \multicolumn{3}{c}{Episode 2} \\
\cline{2-4}\cline{6-8}
\colhead{} & \colhead{$\mu$} & \colhead{$\sigma$} & \colhead{$\chi_\nu^2$} & \colhead{} & \colhead{$\mu$} & \colhead{$\sigma$} & \colhead{$\chi_\nu^2$} 
}
\startdata
$\tau_{\rm bb}^{*}$ (ms) & $121^{+45}_{-33}$ & $0.52\pm0.14$ & 0.38 &  &$182^{+22}_{-19}$ & $0.52\pm0.04$ & 1.29 \\
\cline{1-8}
BB+BB~$kT_{\rm low}$~(keV) & $4.0\pm0.7$ & $1.0\pm0.7$ & 1.00 & & $4.5\pm0.1$ & $1.0\pm0.1$ & 1.17 \\
BB+BB~$kT_{\rm high}$~(keV) & $13.6\pm1.3$ & $2.6\pm1.7$ & 0.91 & & $9.4\pm2.8$ & $4.3\pm1.7$ & 1.6 \\
\cline{1-8}
COMPT~$E_{\rm peak}$~(keV) & $27.0\pm1.0$ & $2.4\pm0.8$ & 1.00 & & $26.3\pm0.7$ & $4.3\pm0.6$ & 1.69 \\
COMPT~$\Gamma$ & $-0.31\pm0.89$ & $0.89\pm1.07$ & 0.04 & & $-0.10\pm0.12$ & $0.67\pm0.12$ & 0.13  
\enddata
\tablecomments{$^*$ $\sigma$ is in the logarithmic scale.}
\end{deluxetable*}

Another measure of a burst duration, is $T_{90}$, that is the time interval over which the cumulative energy fluence of the burst increases from 5\% to 95\% of the total  \citep{Kouveliotou1993}. 
\citet{Lin2020} showed that the \tbb is tightly correlated with $T_{90}$ for \sgr bursts. Note that \tbb is slightly longer, as it measures the full duration of the event while $T_{90}$ measures 10\% less, to account for background fluctuations preceding and following an event. Here we only report \tbb durations. 

\subsection{Spectral Analysis} \label{sec:spectral}

For each burst, we identified the NaI detectors with a $\leq$50$^{\circ}$ angle between the detector zenith and the source at the time of the burst, and also not blocked by other parts of the spacecraft (using \textit{gbmblock}). We then generated response matrices for each detector using the position of \sgr at the start time of each burst with the \textit{gbmrsp} software. We performed spectral modelling with the RMFIT suite, using Cash statistics \citep{cash79}.

The time-integrated burst spectra were extracted using the \tbb interval and were fit with two continuum models which represent magnetar burst spectra the best: the sum of two blackbody functions (BB+BB) and the Comptonized model (COMPT)\footnote{The Comptonized model is an exponentially cutoff power law with the photon number flux ${\cal F} \propto E^{\Gamma}{\rm exp}[-E(2+\Gamma)/E_{\rm peak}]$, where $E_{\rm peak}$ is the peak energy and $\Gamma$ is the photon index.}. Three other simpler models were also fit to the data when neither the BB+BB nor the COMPT model parameters could be well constrained  \citep{Lin2020}. These were: power law (PL), optically thin thermal bremsstrahlung (OTTB), and single blackbody (BB). 
In Table~\ref{tab:model}, 
we summarize the performance of these models in fitting the \sgr burst spectra. 
In Table~\ref{tab:burstlist}, we tabulated the best fit model parameters, fit statistics and their fluence ($8-200$~keV).

\begin{deluxetable*}{lccccccc}
\tablenum{4}
\tablecaption{Continuum models employed in fitting \sgr burst spectra. \label{tab:model}}
\tablewidth{0pt}
\tablehead{
\colhead{Episode} & \colhead{Number of bursts} & \colhead{BB+BB\textcolor{red}{$^{\dagger}$} (\%)} & \colhead{COMPT\textcolor{red}{$^{\ddagger}$} (\%)} & \colhead{Both\textcolor{red}{$^{*}$}} & \multicolumn{3}{c}{Simple models\textcolor{red}{$^{**}$}} \\
\cline{6-8}
\colhead{} & \colhead{} & \colhead{} & \colhead{} & \colhead{} & \colhead{OTTB} & \colhead{PL} & \colhead{BB}
}
\startdata
1 & 21 & 13 (62) & 7 (33) & 6 & 3 & 1 & 3 \\
2 & 125 & 76 (61) & 48 (39) & 44 & 31 & 9 & 5\\
all & 148 & 90 (61) & 56 (38) & 51 & 35 & 10 & 8
\enddata
\tablecomments{
$^{\dagger}$ The number and percentage of bursts that can be adequately fit with the BB+BB model. \\
$^{\ddagger}$ The number and percentage of bursts that can be adequately fit with the COMPT model. \\
$^*$ The number of bursts that can be fit with both BB+BB and COMPT models. \\
$^{**}$ The number of bursts that can only be fit with simple models (OTTB, PL or BB). 
}
\end{deluxetable*}

\begin{figure*}
\includegraphics*[viewport=75 165 600 495, scale=0.5]{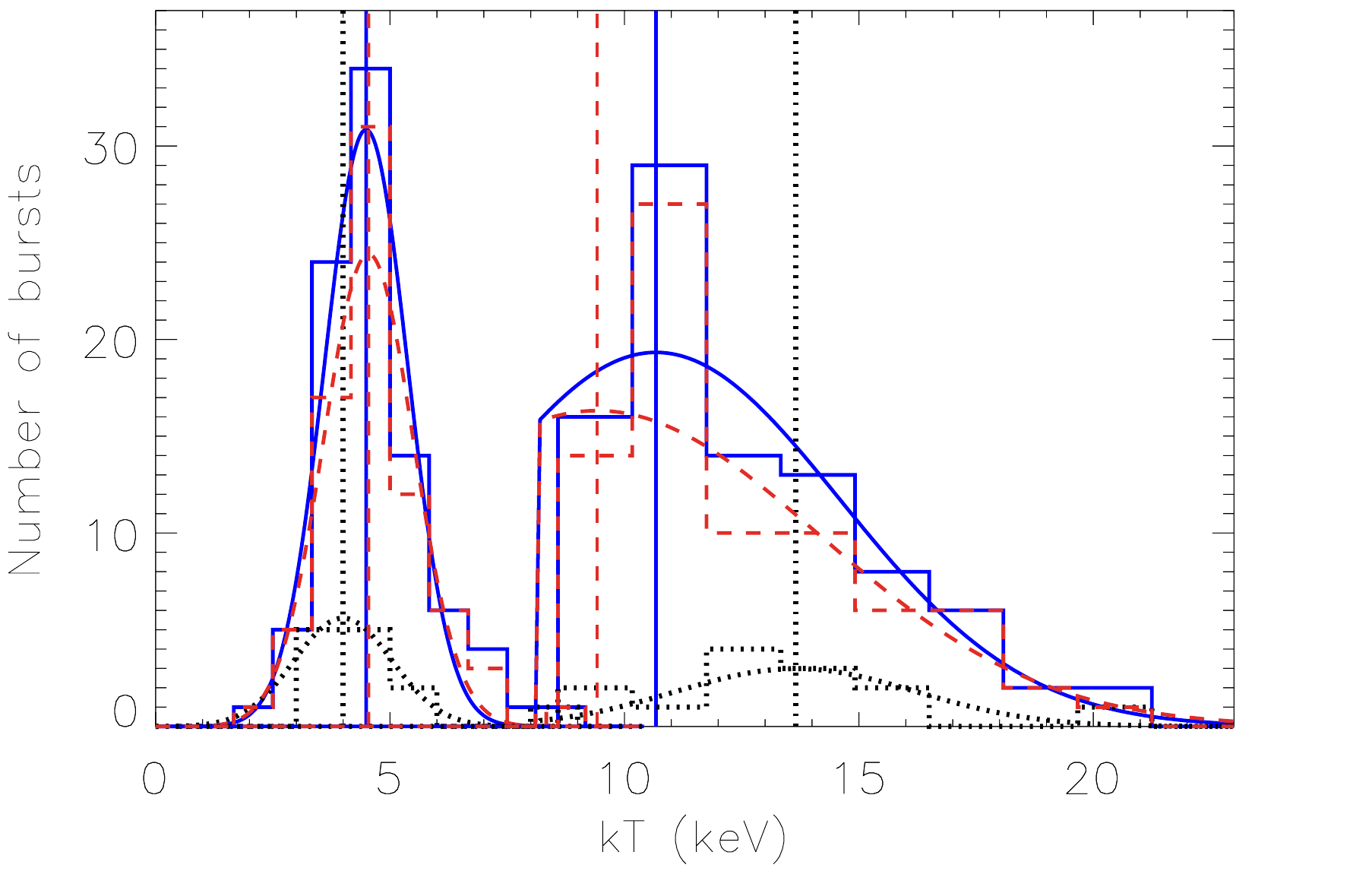}
\includegraphics*[viewport=30 170 600 510, scale=0.5]{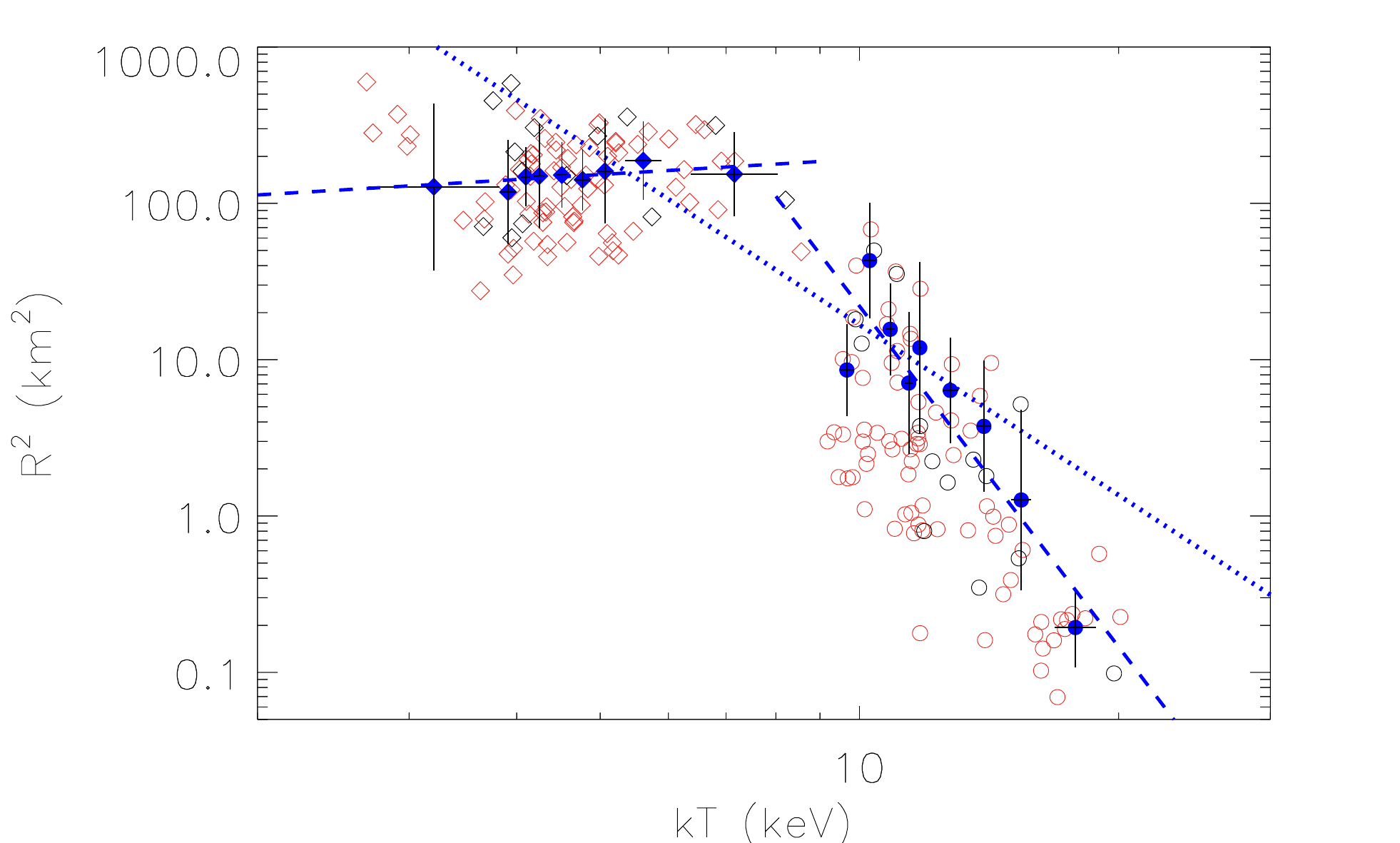}
\caption{\textit{Left:} The distributions of the low and high BB temperatures derived with the BB+BB model with the best-fit Gaussian curves and their mean values. The blue-solid lines, black-dotted lines and red-dashed lines represent all bursts, bursts in 2019, and in 2020, respectively.
\textit{Right:} The emission areas ($R^2$) as a function of the low  (diamonds) and high  (circles) BB temperatures. The blue-filled symbols represent values for groups of every ten data points. The blue-dashed lines indicate the PL fit to the grouped data of each BB temperature emission region. The blue-dotted line is the PL fit of both BB components in linear $R^2-T$ space. The colors denote events in episodes 2 (red), and 1 (black).
\label{fig:specBBBB}}
\end{figure*}

The left panel of Figure~\ref{fig:specBBBB} shows the distributions of both the low and high BB temperatures for the 90 bursts that were adequately represented with the BB+BB model.
The low BB temperature follows a Gaussian trend with the best fit mean value of 4.5$\pm$0.1~keV. The distribution of the high BB temperature is asymmetric due to its overlap with the low BB component and is best fit with a truncated Gaussian function with a lower cutoff at the highest low BB temperature (8.2~keV), resulting in a mean value of 10.7$\pm$1.3~keV.
We also note here that when similar analyses were performed individually for the two burst episodes, their temperatures agreed within statistical errors, as shown in Table~\ref{tab:distribution}. 

Next, we investigated how the best-fit model parameters and the calculated fluences correlated with each other. 
We present these correlations in Table \ref{tab:correlation} with the results of their power law fits obtained from linear fits in logarithmic scale, as well as the parameters of each Spearman's rank order correlation test.\footnote{We caution that artifacts may affect the results when subdividing into the low and high temperature BB components.} 
We find the size of the BB emitting regions ($R^{2}$) and energy fluence ($F$ and thus luminosities) of both BB components to be strongly correlated for the 90 bursts in our sample ($R_{high}^2 \propto (R_{low}^2)^{\alpha}$, $F_{high} \propto F_{low}^{\alpha}$ and $L_{high} \propto L_{low}^{\alpha}$ in Table \ref{tab:correlation}), as are the areas ($R^{2}$) and the temperatures of the two BB components ($R^2 \propto kT^{\alpha}$ in Table \ref{tab:correlation}).
The high BB temperature component was found to be inversely proportional to the emission area (the right panel of Figure~\ref{fig:specBBBB}).
In contrast, the emission area of the low temperature BB component is relatively constant across its entire temperature range.
There is significant scatter in the temperatures and emission areas for both BB components in the ensembles: a power law fit to the $R^2-T$ correlation may be highly affected by a few outliers.  Accordingly, we grouped every ten data points and performed the PL fit for each BB component on the grouped data as illustrated in the right panel of Figure~\ref{fig:specBBBB}.  The fit results are listed in Table~\ref{tab:correlation}. 
Interestingly, the emission area dependence spanning both the low and high BB temperatures, $R^2 \propto (kT)^{-3.6\pm0.2}$, was very similar to the one corresponding to a single BB obeying the Stefan-Boltzmann law: $R^2 \propto (kT)^{-4}$.  This $R-T$ correlation for BB+BB fits is also very close to that observed for the collection of SGR~J1550$-$5418 bursts analyzed in the studies of \cite{lin2012} and \cite{AvdH2012}.  It is evident that for the entire BB+BB fitting ensemble, $R^2 T^4$ is an increasing function of $T$ and hence also burst flux.  Thus, brighter bursts are on average slightly harder in their BB+BB fits, noting that the same weak flux-hardness correlation is identified just below for the bursts with preferred COMPT fits.


\begin{deluxetable*}{cccc}
\tablenum{5}
\tablecaption{Results of Spearman test and power law fit to parameter correlations. \label{tab:correlation}}
\tablewidth{0pt}
\tablehead{
\colhead{Correlation$^{\dagger}$} & \colhead{PL fit index} & \multicolumn{2}{c}{Spearman test} \\
\cline{3-4}
\colhead{} & \colhead{$\alpha$} & \colhead{correlation coefficient} & \colhead{chance probability}
}
\startdata
$R_{high}^2 \propto (R_{low}^2)^{\alpha}$ & $3.1\pm0.7$ & 0.6 & $3.5\times10^{-12}$ \\
$F_{high} \propto F_{low}^{\alpha}$ & $1.1\pm0.1$ & 1.0 & 0 \\
$L_{high} \propto L_{low}^{\alpha}$ & $1.2\pm0.1$ & 0.9 & $5.6\times10^{-45}$ \\
$R^2 \propto kT^{\alpha}$ & $-3.6\pm0.2$ & $-0.8$ & 0 \\
$R_{high}^2 \propto kT_{high}^{\alpha}$ & $-7.2\pm1.3^{*}$ & -0.7 & $7.4\times10^{-13}$ \\
$R_{low}^2 \propto kT_{low}^{\alpha}$ & $0.3\pm1.1^{*}$ & -0.01 & 0.95 \\
\hline
$E_{\rm peak} \propto F^{\alpha}$ & $0.09\pm0.003^{*}$ & 0.6 & $8.2\times10^{-6}$ \\
$\Gamma~v.s.~F$ & \nodata & 0.4 & $1.5\times10^{-3}$ 
\enddata
\tablecomments{
$^{\dagger}$ $R^2$, $F$, $L$ and $kT$ are the emitting area, fluence, luminosity and temperature of a BB, respectively. \\
$^*$ Power law fit to the grouped data. 
}
\end{deluxetable*}

\begin{figure*}
\includegraphics*[viewport=50 140 620 500, scale=0.45]{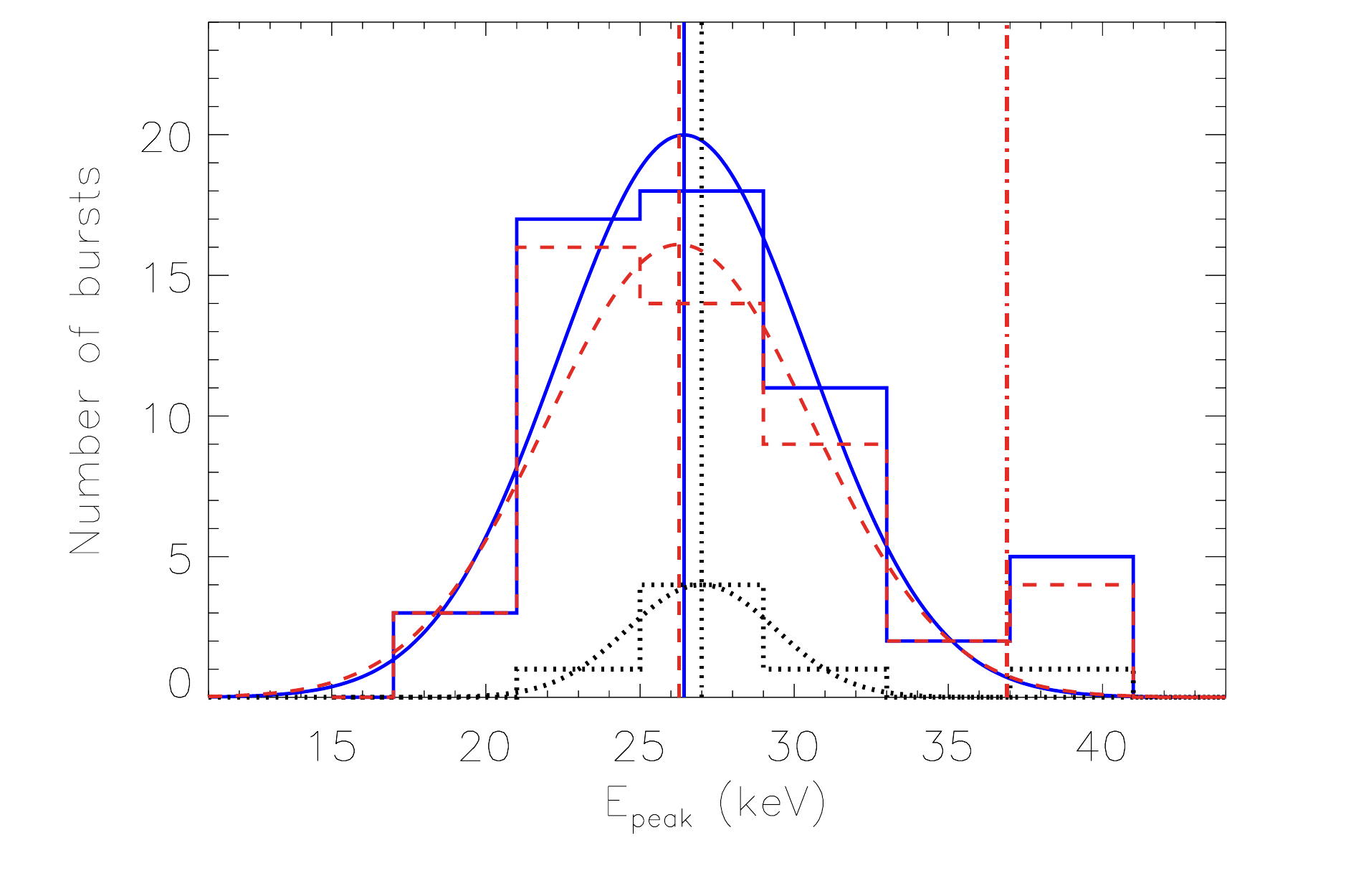}
\includegraphics*[viewport=70 140 620 500, scale=0.45]{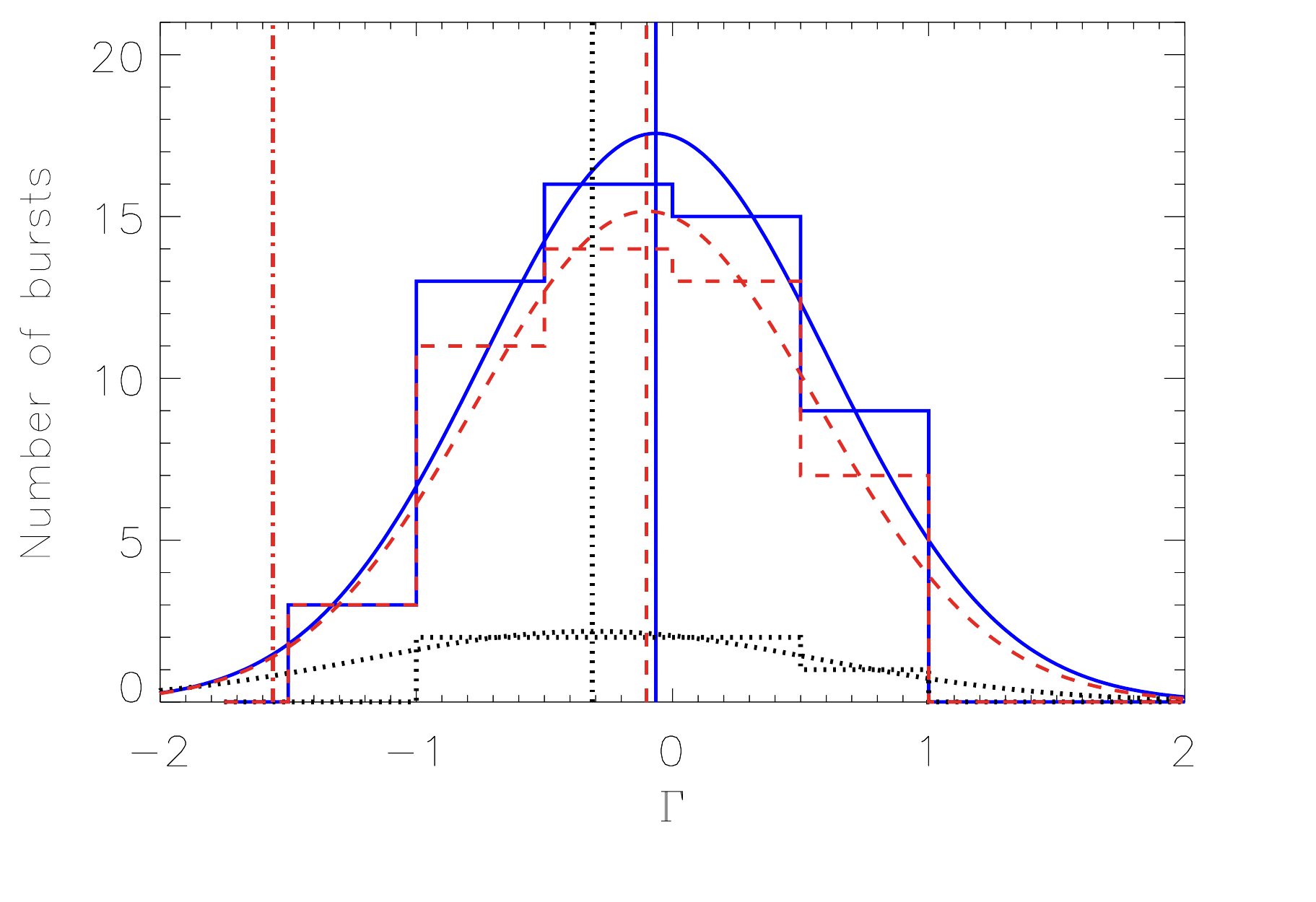}
\includegraphics*[viewport=50 140 620 500, scale=0.45]{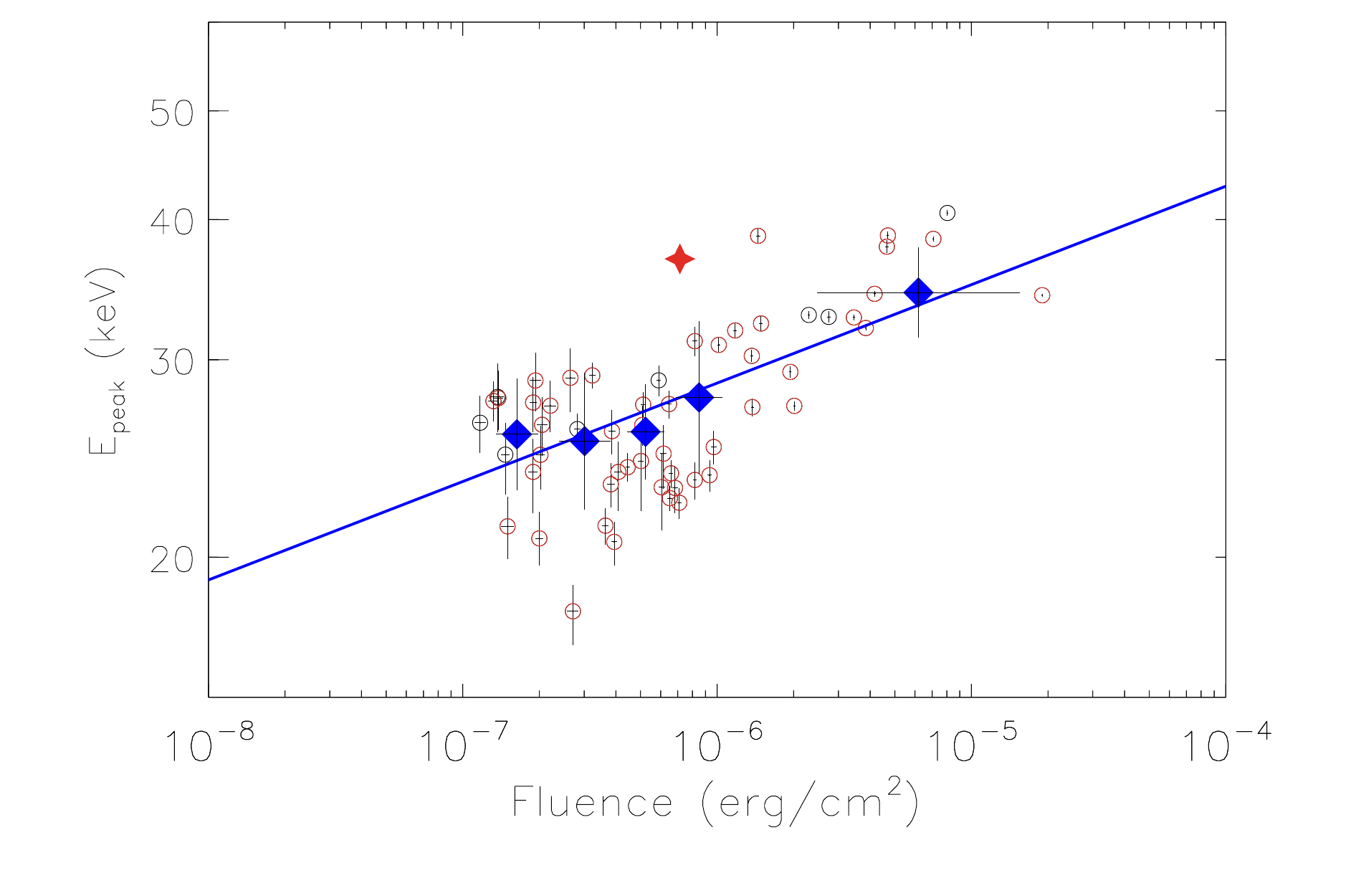}
\includegraphics*[viewport=50 140 620 500, scale=0.45]{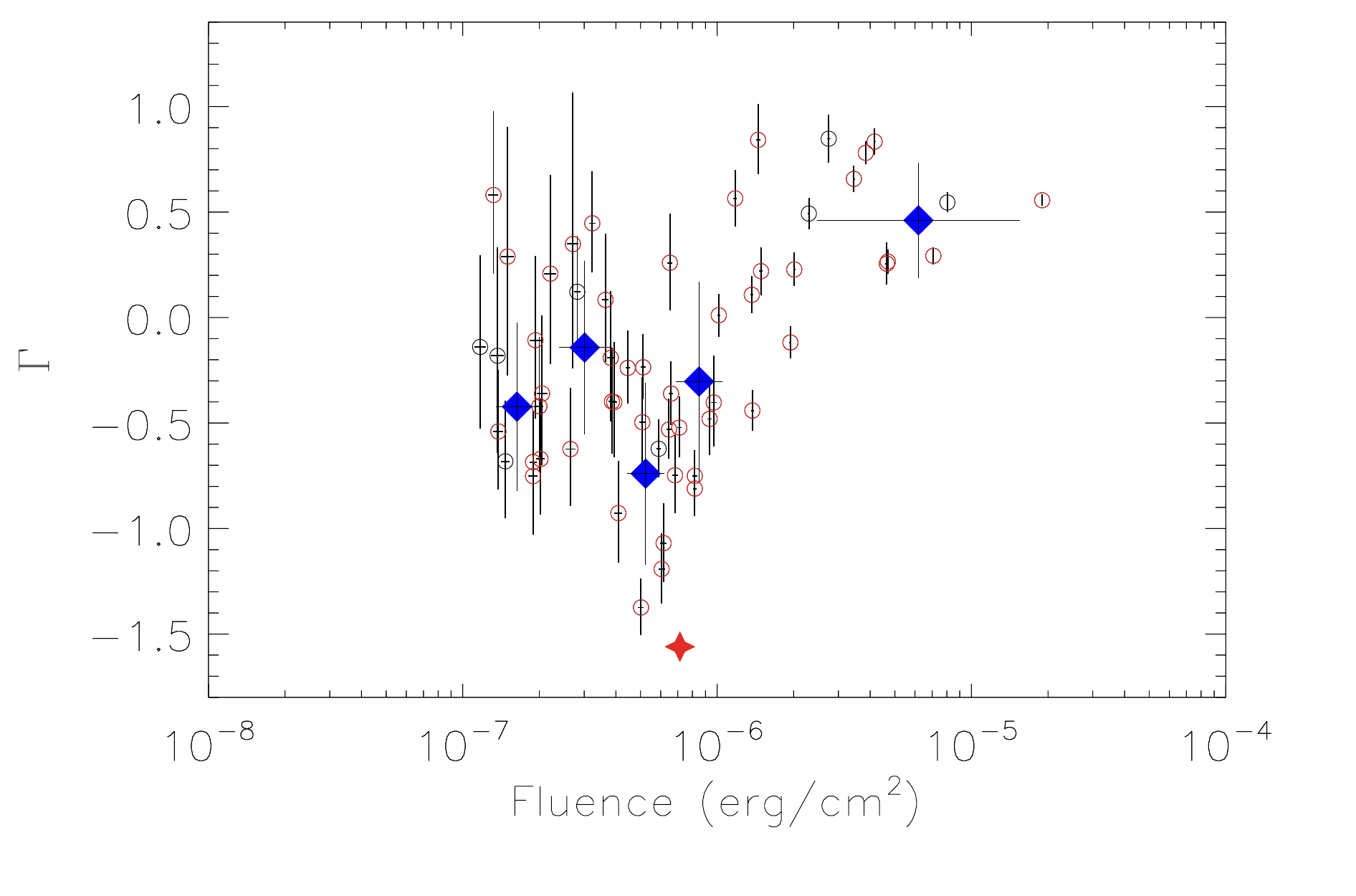}
\caption{The top panels present distributions of $E_{\rm peak}$ (\textit{left}) and $\Gamma$ (\textit{right}) of the COMPT model fits for all bursts (blue solid lines) and bursts in episodes 1 \& 2 (black-dotted and red-dashed lines, respectively). The curves are Gaussian fits to the histograms; their mean values are represented by vertical lines. The lower panels show the $E_{\rm peak}$ (\textit{left}) and $\Gamma$ (\textit{right}) as a function of fluence for all bursts. The bursts in episode 2 are highlighted with red circles. The blue dots represent the weighted means of consecutive groups of ten data points each. The solid line is the best PL fit to the correlation between $E_{\rm peak}$ and fluence. The position of FRB~200428 is shown as a vertical dashed-dotted line in the top panels and as a red star in the bottom panels. \label{fig:specCOMPT}}
\end{figure*}

The COMPT model fits 56 burst spectra well in our sample; seven bursts in the first episode and 48 in the second. Their parameter distributions and correlations are shown in Figure~\ref{fig:specCOMPT}. We find the burst peak energy ($E_{\rm peak}$) to range from 10 to 40~keV, with an average value of $26.4\pm0.6$~keV (derived with a Gaussian fit). The bottom left panel of Figure~\ref{fig:specCOMPT} shows the correlation of $E_{\rm peak}$ with fluence; here we display a weighted average of every ten data points starting from the lowest fluence value due to the large scattering of the data. We clearly observe a positive correlation, indicating that the spectrum becomes harder as the burst fluence increases. The photon index ($\Gamma$) of the COMPT model also follows a Gaussian distribution, with a mean of $-0.06\pm0.12$, over a range of $-1.5$ to $-1.0$. The bottom right panel of Figure~\ref{fig:specCOMPT} shows a weak correlation between $\Gamma$ with burst fluence. We list the quantitative details of these correlations in Table~\ref{tab:correlation}. 

\section{Discussion}

After about three years of quiescence, \sgr has entered another state of heightened burst activity, making it the most prolific transient magnetar. Remarkably, the number of bursts from the 2019 and 2020 episodes in this study, outnumber the total number of all previous bursts since its discovery, without even including the bursts emitted during the burst forest interval.
We discuss below several interesting and somewhat intriguing characteristics from the source's new burst active episodes. 

\begin{figure*}
\includegraphics*[viewport=10 170 600 520, scale=0.45]{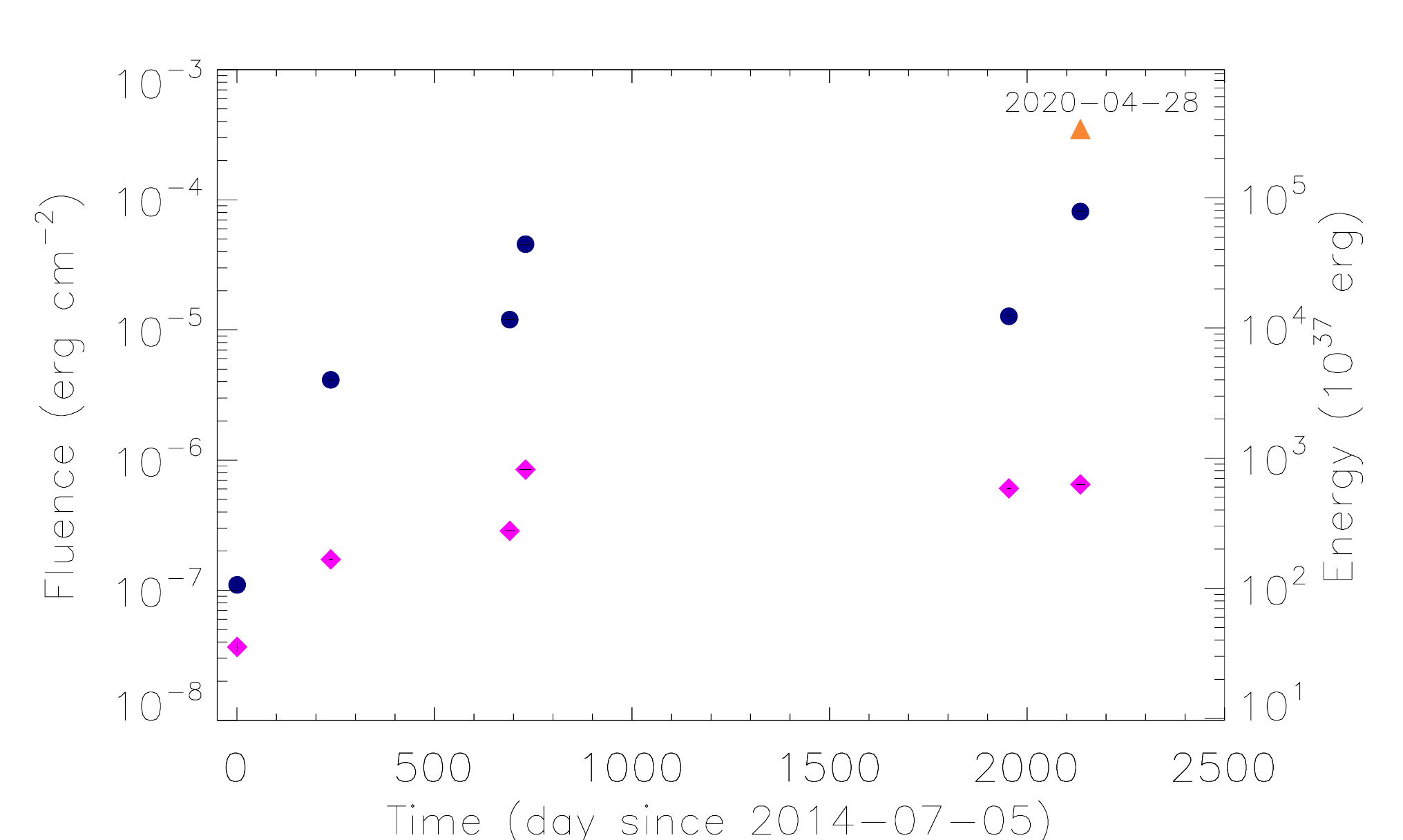} 
\includegraphics*[viewport=60 170 600 500, scale=0.45]{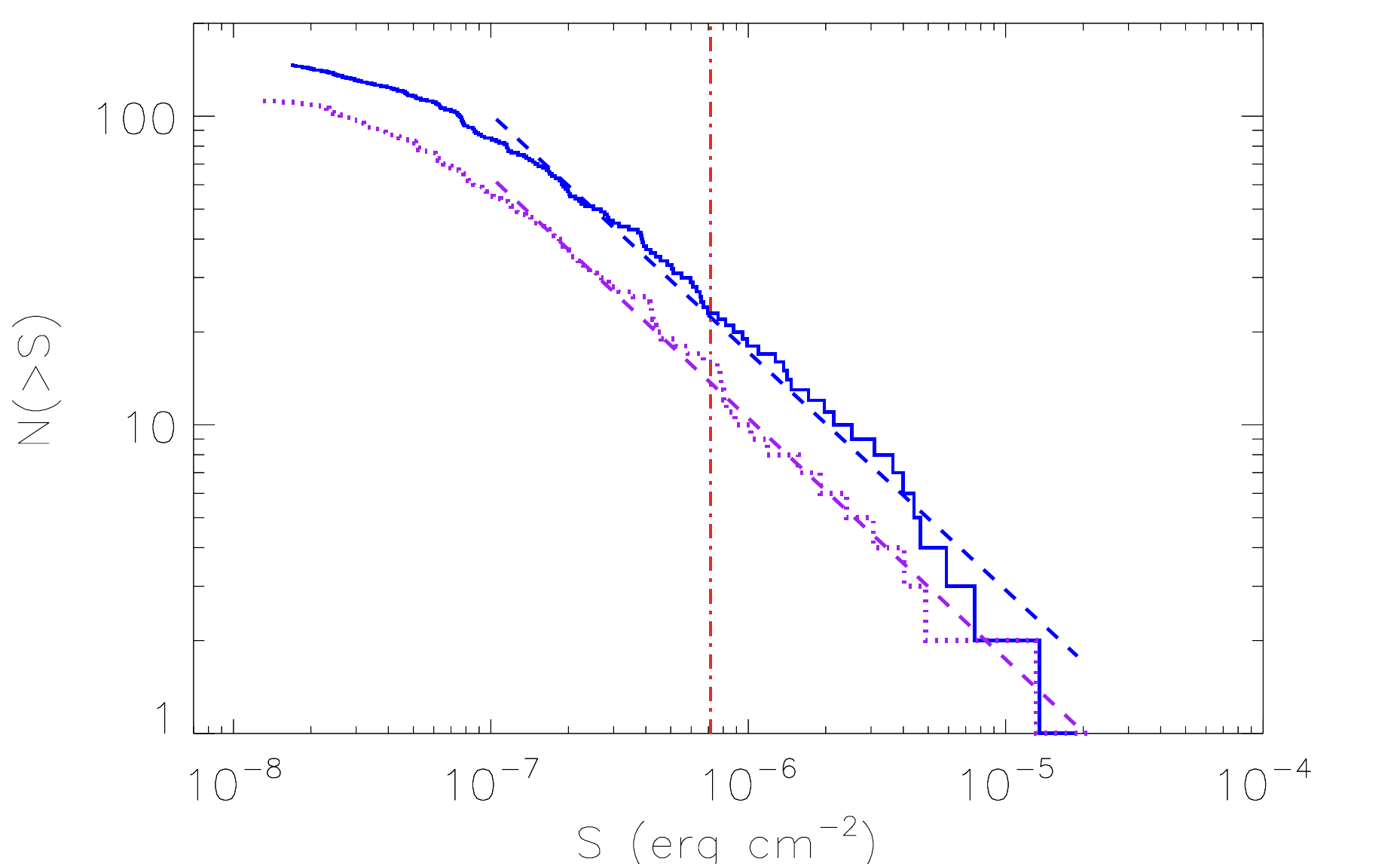}
\caption{\textit{Left:} Time evolution of the total burst fluence (navy dots) and the average fluence per bursts (magenta diamonds from \sgr from discovery to present (left y axis). The orange triangle is the total burst fluence including the burst forest on April 28. The corresponding burst energy, assuming a distance of 9~kpc, is shown in the right y axis. 
\textit{Right:} The cumulative energy fluence distributions of \sgr bursts in 2019-2020 (blue-solid line) and 2014-2016 (purple-dotted line). The two dashed lines are the best PL fit to the distribution above $1\times10^{-7}~\rm{erg}~\rm{cm^{-2}}$. The red vertical dashed-dotted line marks the fluence of the X-ray burst associated with FRB 200428. 
\label{fig:evoflnc}}
\end{figure*}

We present in the left panel of Figure~\ref{fig:evoflnc}, the temporal evolution of the total burst fluence in all burst active episodes since the discovery of \sgrnos, as well as that of the average burst fluence (fluence per burst); both clearly show positive trends.
\citet{Lin2020} reported that the average burst energies (for a distance of 9 kpc) in its 2014, 2015, May 2016 and June 2016 activity episodes were 0.4$\times10^{39}$, 1.7$\times10^{39}$, 2.8$\times10^{39}$ and 8.2$\times10^{39}$ erg, respectively. 
This trend was suggestive of a future higher burst activity; contrary to this expectation, the average burst energies of the 2019 and 2020 episodes, of 5.9$\times10^{39}$ and 6.3$\times10^{39}$ erg, respectively, indicate a flattening of the average burst energy curve. However, these values correspond only to the 148 bursts studied here - adding the contribution of the burst forest in the 2020 episode  significantly increases its final value (see the left panel of Figure~\ref{fig:evoflnc}). We consider, therefore, the current values as lower limits of the source energetics. This also takes into account the bursts that were missed when GBM was occulted by the Earth or in the South Atlantic Anomaly. 

The distribution of the cumulative energy fluence for all 148 bursts from \sgr is shown in the right panel of Figure~\ref{fig:evoflnc}. This distribution is optimally represented with a broken PL, with indices of $0.31\pm0.01$ and $0.72\pm0.02$ for the lower and higher fluences, respectively. The break in the fluence occurs at $1.2\pm0.1\times10^{-7}$~erg~cm$^{-2}$. A single PL model also fits fluences above $S=1\times10^{-7}$ erg cm$^{-2}$, which has generally been used in previous studies as the threshold for the 100\% detection rate 
\citep{AvdH2012,Collazzi2015}. The distribution of bursts with fluences $\geq1\times10^{-7}$ erg cm$^{-2}$ is well fit with a PL, with an index of -0.77$\pm$0.01. This is very consistent with the PL index of -0.78 for the cumulative burst fluence in previous active episodes from this source \citep{Lin2020}. It is important to note that although the 2019 and 2020 bursts were more energetic on average, they follow the same trend with past activations, as shown in the right panel of Figure~\ref{fig:evoflnc}.


The spectroscopy of the bursts provides information on the physical
environment, where their emission originated.  In general, by setting $E_{\rm peak}\sim 3kT_{e,max}$, one obtains an estimate of the maximum for 
the effective plasma temperature $T_e$ in the inner magnetospheric emission
region.  The $\tbb$ values vastly exceed the typical dynamical times
$\rns/c\sim 3\times 10^{-5}$~s for a neutron star radius $\rns\sim 10^6$~cm, 
so that plasma is nominally trapped in closed 
magnetic field line regions that are somewhat remote from the magnetic poles.
Sub-surface crustal dislocation by the strong fields likely leads to the energy deposition in the magnetosphere \citep{td95}, heating the pair plasma.
With such an injection from the surface, effective temperature gradients are 
likely to be established due to the adiabatic cooling of gas as it expands to
high altitudes.  The convolution of such gradients will present itself as 
somewhat similar to the apparently non-thermal spectra in the data,
masquerading as BB+BB or COMPT forms.  The 
energetics of bursts guarantees optically thick plasma with highly saturated, 
Comptonized spectra at each magnetospheric locale, as discussed in
\cite{lin11,lin2012}.  Within the total (putatively quasi-equatorial) emission region,
energy conservation for the plasma+radiation transport from one zone to another 
connected by magnetic flux tubes dictates that when approaching thermal equilibrium, 
though not fully realizing it, the Stefan-Boltzmann law $R^2T_e^4=$constant 
is approximately satisfied.  This is the physical origin of the observed
high/low temperature coupling in the BB+BB fits.

Yet the BB fitting protocol does not automatically imply an absolutely thermal emission region. One can estimate the average
detected flux ${\cal F}$ for each burst in Episode 2 using the total
accumulated fluence listed in Table \ref{tab:burstepisode} divided by the number of bursts (125), further divided by
$kT_{high}\sim 15$keV and also by the average $\tbb \sim 200\;$ms identified in Figure~\ref{fig:tbb}. From this, one can compute the
photon number density $n_{\gamma}\sim {\cal F}d^2/(R^2c)$ typically
expected in the magnetospheric emission region.  Assuming a source
distance of $d=9$ kpc and an emission region size of $R=10^6\,$cm, one
arrives at $n_{\gamma}\sim 3 \times 10^{24}$cm$^{-3}$.  This is
considerably smaller than the density $0.24\, (kT_{high}/[\lambdaC\,
m_ec^2])^3 \sim 10^{26}$cm$^{-3}$ of a pure Planck distribution of
temperature $T_{high}$, for a reduced electron Compton wavelength
$\lambdaC = \hbar/m_ec$. It is thus anticipated that thermalization is locally
significant, though incomplete.

\begin{figure*}
\includegraphics*[viewport=30 160 600 510, scale=0.45]{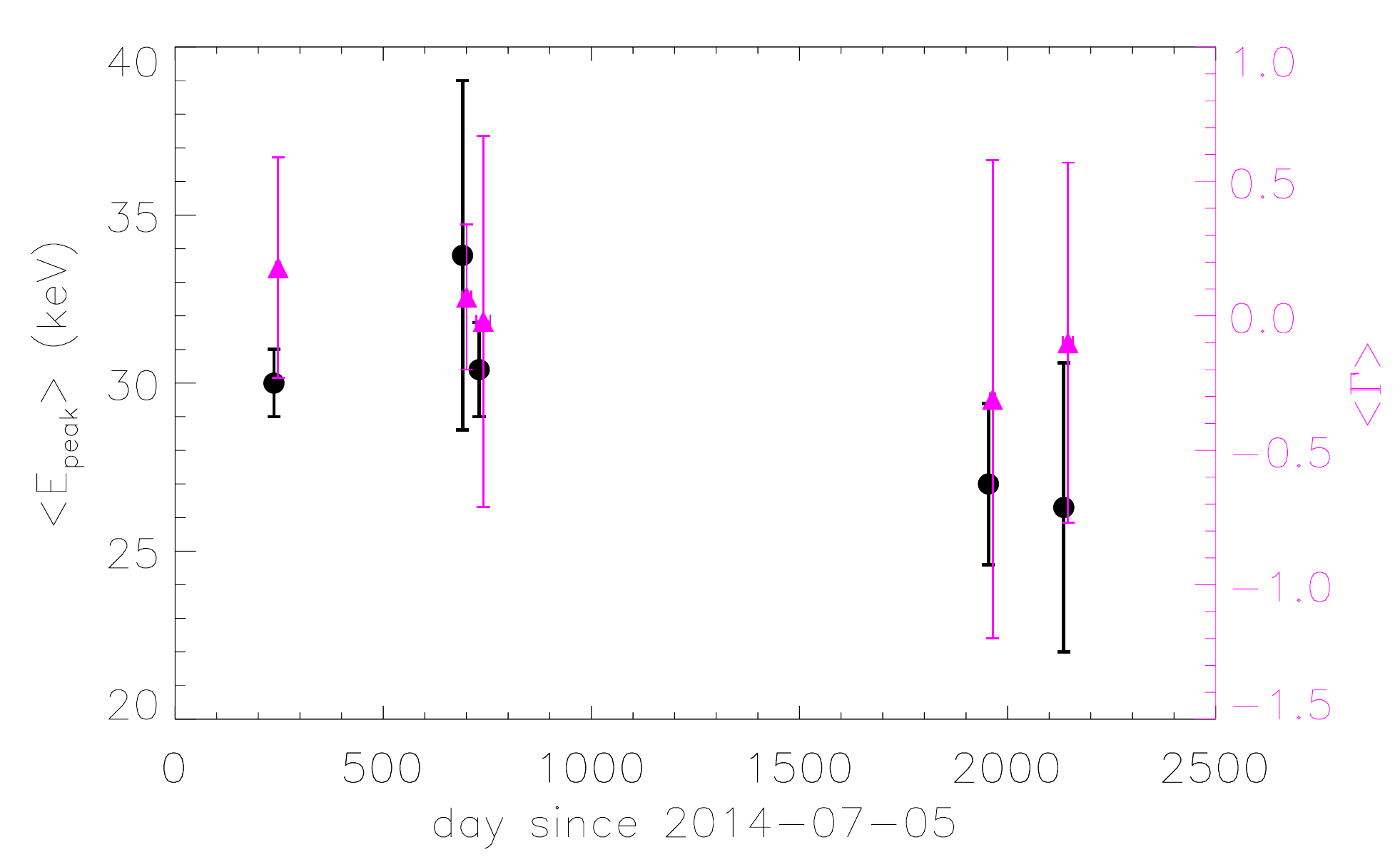}
\includegraphics*[viewport=10 160 600 510, scale=0.45]{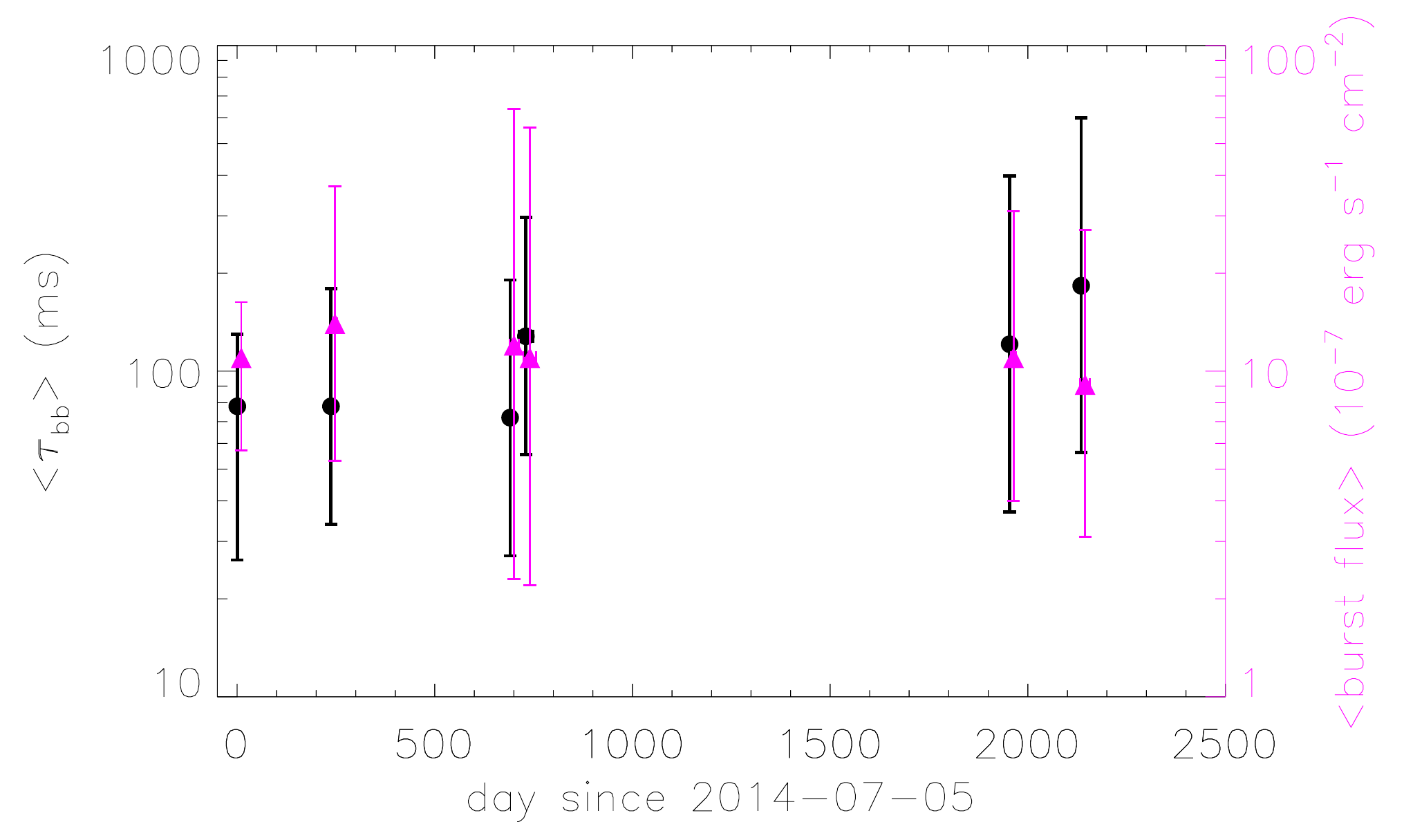}
\caption{The evolution of $E_{\rm peak}$ (\textit{left}, black dots), $\Gamma$ (\textit{left}, magenta triangles), \tbb (\textit{right}, black dots) and averaged burst flux (\textit{right}, magenta triangles) in $8-200$~keV throughout the six years of activity from \sgr (2014 to 2020). The error bars are the $1\sigma$ standard deviation of the Gaussian distribution. The magenta triangles are shifted to right by 10 days for better visibility.}  \label{fig:evoepeak}
\end{figure*}

The comparison of the average
\footnote{It is the mean value of the Gaussian fit to the distribution of $E_{\rm peak}$. This is also the case for average $\Gamma$ and $\tbb$.}
$E_{\rm peak}$ of bursts between 2014 to 2016 indicates a slight drop in hardness when progressing from that epoch to the 2019/2020 bursts in this study, although this variation is within the one sigma level: $E_{\rm peak}$ drops from $\sim30-35$~keV to $~27$~keV, respectively (the left panel of Figure~\ref{fig:evoepeak}). Combining this trend with the rise in fluence exhibited in the left panel of Figure~\ref{fig:evoflnc} over the same period suggests an anti-correlation between the average $E_{\rm peak}$ and fluence. Note that this is {\it opposite} to the $E_{\rm peak}-F$ trend in Figure~\ref{fig:specCOMPT} present for the 
2019-2020 burst population.  This evolutionary character is underpinned by an increase in
the average 
burst duration $\tbb$ for the 2019-2020 bursts relative to the historic ones: see the right panel of Figure~\ref{fig:evoepeak}.
We note that short bursts from other magnetars typically have an $E_{\rm peak}$ of $\sim40$~keV \citep{Collazzi2015}, indicating that bursts from \sgr are also somewhat softer, corresponding to cooler plasma temperatures.  Yet, noting the trend of increasing burst fluence over the 2014--2020 period, it is plausible to 
assume that the energy deposited into the magnetosphere (about 10$^{39}$ erg) to precipitate these bursts is actually slightly increasing over this 6-year interval.
Given that the sizes of the emitting area for the high temperature BB 
component in our sample are consistent with that of other magnetars 
\citep{AvdH2012}, we propose that the cooling of the maximum effective plasma temperature of \sgr bursts over time could correspond to greater masses and 
densities in the magnetospheric plasma emitting the bursts on average, 
and hence higher opacities.  The likely coupling between such densities,
temperature and the spectral index as discussed in \cite{lin11,lin2012} can help provide diagnostics for models of polarized radiative transport that lead to the generation of the spectra studied here.

A non-thermal spectrum has been reported for the hard X-ray burst associated with FRB~200428 from \sgr, 
with parameters $\Gamma\sim-1.5$ and $E_{\rm peak} \sim 37$~keV when converted to our presentation here of the COMPT model \citep{li2020}.
This peak energy is slightly higher than that of bursts with similar fluences in our sample (see the lower-left panel of Figure~\ref{fig:specCOMPT}). Therefore, the X-ray burst associated with the FRB is a slightly harder magnetar burst, yet with a noticeably steeper spectrum, a contrast highlighted in \citet{younes20_2}. As discussed above, this peculiar burst might have originated from a low density plasma region. 
Indeed the PL index of the burst associated with FRB~200428, as reported by \citet{li2020}, is the steepest (softest) compared to the earlier bursts from \sgr or bursts from other magnetars. 
This is in agreement with the joint spectral analysis of GBM and NICER for \sgr \citep{younes20_2} and GBM and Swift/XRT data for SGR~J1550$-$5418 \citep{lin2012}. In order to reach a typical $E_{\rm peak}$ with a soft index, the overall spectral curvature needs to be rather flat, close to a power law with a relatively high cutoff energy \citep{li2020,kw200428}. The 56 bursts in our sample that can be fit with the COMPT model reveal a softer $E_{\rm peak}$ with a typically harder photon index. This suggests a larger curvature in the spectral shape, indicating a more thermalized spectrum. This is also in agreement with the previous broadband spectral analysis of other magnetar bursts \citep{israel2008,lin2012}. A more thermalized spectrum may indicate an environment with a higher plasma density and thus scattering opacity, with the emission region perhaps spanning smaller ranges of magnetospheric altitudes.  High opacity is extremely destructive for coherent radio emission mechanisms, and so it is reasonable to assert that radio signals are less likely to be generated in association with these putatively higher density bursts. This is in agreement with the non-radio detection of radio pulses from other \sgr bursts \citep{Lin2020b}.

Recently three faint FRB-like events from \sgr were detected, one on April $30^{th}$ 2020 \citep{fastatel2020b} and two on May $24^{th}$ 2020 \citep{2020arXiv200705101K}. At the time of the first radio burst, the GBM line of sight to the magnetar was occulted by the Earth. The times of the latter two events, which were separated by only 1.4 s from each other, were within the GBM field of view and their time span was covered by our search for untriggered events;
we did not find any X-ray bursts coincident with these radio bursts. We place a 3$\sigma$ flux upper limit in the 8-200 keV band of 2.2$\times$10$^{-8}$ erg cm$^{-2}$ s$^{-1}$, assuming bursts with 0.5~s duration and with the same spectral shape with that of the burst associated with FRB~200428. This further implies that the flux ratio between X-ray and the May $24^{th}$ 2020 radio events is less than $10^{-9}$~(erg cm$^{-2}$)/(Jy~ms).


\clearpage

\begin{longrotatetable}
\begin{deluxetable*}{lllcllllll}
\tablecolumns{10}
\tablenum{1}
\tablecaption{The \sgr burst list. \label{tab:burstlist}}
\tablewidth{0pt}
\tablehead{
\colhead{ID} & \colhead{Burst start time} & \colhead{T$_{\rm bb}$} & \colhead{$F^{a}$} &  \colhead{E$_{\rm peak}$} & \colhead{$\Gamma$} & \colhead{C-Stat/DoF~$^{b}$}  & \colhead{kT$_{\rm low}$} & \colhead{kT$_{\rm high}$} & \colhead{C-Stat/DoF~$^{c}$}\\
\colhead{} & \colhead{in UTC} & \colhead{(s)} & \colhead{$( 10^{-7}~\rm{erg~cm}^{-2})$} & \colhead{(keV)} & \colhead{} & \colhead{} & \colhead{(keV)} & \colhead{(keV)} & \colhead{} 
}  
\startdata
\multicolumn{10}{c}{2019} \\
\cline{1-10}
1	$^{T}$	& Oct 04 09:00:53.609	&	0.095	&$	0.74 	\pm	0.05 	$&	$	26.48 	_{-	2.84 	}^{+	3.21 	}$	&	\nodata							&	$	207.96 	/	201 	$	&	\nodata							&	\nodata							&	\nodata					\\
2	$^{T}$	& Nov 04 01:20:24.034	&	0.092	&$	1.05 	\pm	0.06 	$&	\nodata							&	\nodata							&	\nodata					&	$	4.06 	_{-	0.48 	}^{+	0.49 	}$	&	$	12.66 	_{-	1.39 	}^{+	1.67 	}$	&	$	216.52 	/	198 	$	\\
3	$^{T}$	& Nov 04 02:53:31.369	&	0.035	&$	1.47 	\pm	0.06 	$&	$	24.69 	_{-	1.94 	}^{+	1.68 	}$	&	$	-0.68 	_{-	0.27 	}^{+	0.29 	}$	&	$	286.03 	/	270 	$	&	$	3.94 	_{-	0.75 	}^{+	0.72 	}$	&	$	9.90 	_{-	0.93 	}^{+	1.23 	}$	&	$	289.27 	/	269 	$	\\
4	$^{T}$	& Nov 04 04:26:55.855	&	0.074	&$	1.17 	\pm	0.06 	$&	$	26.36 	_{-	1.58 	}^{+	1.51 	}$	&	$	-0.14 	_{-	0.39 	}^{+	0.44 	}$	&	$	191.97 	/	199 	$	&	$	5.74 	_{-	0.52 	}^{+	0.45 	}$	&	$	15.31 	_{-	3.30 	}^{+	5.11 	}$	&	$	186.41 	/	198 	$	\\
5		& Nov 04 07:20:33.684	&	0.100	&$	0.48 	\pm	0.05 	$&	$	24.86 	_{-	3.65 	}^{+	4.28 	}$	&	\nodata							&	$	308.87 	/	266 	$	&	\nodata							&	\nodata							&	\nodata					\\
6		& Nov 04 08:56:15.943	&	0.043	&$	0.17 	\pm	0.03 	$&	\nodata							&	\nodata							&	\nodata					&	$	6.11 	_{-	0.73 	}^{+	0.84 	}$	&	\nodata							&	$	220.52 	/	201 	$	\\
7	$^{T}$	& Nov 04 09:17:53.492	&	0.321	&$	5.89 	\pm	0.13 	$&	$	28.76 	_{-	0.93 	}^{+	0.90 	}$	&	$	-0.62 	_{-	0.14 	}^{+	0.14 	}$	&	$	332.12 	/	268 	$	&	$	4.57 	_{-	0.27 	}^{+	0.27 	}$	&	$	11.76 	_{-	0.55 	}^{+	0.62 	}$	&	$	326.49 	/	267 	$	\\
8	$^{T}$	& Nov 04 10:44:26.231	&	0.195	&$	22.94 	\pm	0.23 	$&	$	32.89 	_{-	0.29 	}^{+	0.29 	}$	&	$	0.49 	_{-	0.07 	}^{+	0.08 	}$	&	$	268.16 	/	199 	$	&	$	5.38 	_{-	0.38 	}^{+	0.38 	}$	&	$	10.40 	_{-	0.30 	}^{+	0.35 	}$	&	$	269.50 	/	198 	$	\\
9	$^{T}$	& Nov 04 12:38:38.534	&	0.072	&$	2.82 	\pm	0.09 	$&	$	26.02 	_{-	0.88 	}^{+	0.85 	}$	&	$	0.12 	_{-	0.25 	}^{+	0.27 	}$	&	$	232.05 	/	200 	$	&	$	4.97 	_{-	0.57 	}^{+	0.57 	}$	&	$	10.06 	_{-	1.04 	}^{+	1.62 	}$	&	$	228.29 	/	199 	$	\\
10	$^{T}$	& Nov 04 15:36:47.402	&	0.321	&$	1.23 	\pm	0.08 	$&	\nodata							&	\nodata							&	\nodata					&	$	3.66 	_{-	0.54 	}^{+	0.58 	}$	&	$	11.89 	_{-	1.17 	}^{+	1.37 	}$	&	$	356.40 	/	336 	$	\\
11		& Nov 04 19:09:01.727	&	0.038	&$	0.29 	\pm	0.04 	$&	$	23.85 	_{-	4.60 	}^{+	5.61 	}$	&	\nodata							&	$	120.25 	/	133 	$	&	\nodata							&	\nodata							&	\nodata					\\
12		& Nov 04 20:01:41.871	&	0.127	&$	0.46 	\pm	0.05 	$&	\nodata							&	\nodata							&	\nodata					&	$	3.95 	_{-	0.59 	}^{+	0.65 	}$	&	$	13.77 	_{-	2.23 	}^{+	2.78 	}$	&	$	301.79 	/	264 	$	\\
13		& Nov 04 20:13:42.537	&	0.140	&$	0.61 	\pm	0.07 	$&	$	42.00 	_{-	6.62 	}^{+	8.16 	}$	&	\nodata							&	$	210.03 	/	200 	$	&	\nodata							&	\nodata							&	\nodata					\\
14	$^{T}$	& Nov 04 20:29:39.804	&	0.128	&$	1.37 	\pm	0.08 	$&	$	27.80 	_{-	1.96 	}^{+	1.96 	}$	&	$	-0.18 	_{-	0.46 	}^{+	0.51 	}$	&	$	172.31 	/	133 	$	&	\nodata							&	\nodata							&	\nodata					\\
15		& Nov 04 23:16:49.544	&	0.024	&$	0.25 	\pm	0.03 	$&	\nodata							&	\nodata							&	\nodata					&	$	9.71 	_{-	0.78 	}^{+	0.85 	}$	&	\nodata							&	$	243.89 	/	267 	$	\\
16	$^{T}$	& Nov 04 23:48:01.336	&	0.225	&$	3.07 	\pm	0.11 	$&	\nodata							&	\nodata							&	\nodata					&	$	3.98 	_{-	0.33 	}^{+	0.33 	}$	&	$	12.15 	_{-	0.97 	}^{+	1.10 	}$	&	$	248.05 	/	199 	$	\\
17		& Nov 05 00:33:02.781	&	0.881	&$	1.17	\pm	0.22	$&	\nodata							&	$	-2.293	_{-	0.298	}^{+	0.254	}$	&	$	71.775	/	65 	$	&	\nodata							&	\nodata							&	\nodata					\\
18	$^{T}$	& Nov 05 06:11:08.595	&	0.786	&$	80.42 	\pm	0.57 	$&	$	40.55 	_{-	0.24 	}^{+	0.24 	}$	&	$	0.55 	_{-	0.05 	}^{+	0.05 	}$	&	$	362.53 	/	132 	$	&	$	8.21 	_{-	0.23 	}^{+	0.22 	}$	&	$	15.39 	_{-	0.60 	}^{+	0.68 	}$	&	$	337.39 	/	131 	$	\\
19	$^{T}$	& Nov 05 07:17:17.705	&	0.194	&$	0.91 	\pm	0.08 	$&	\nodata							&	\nodata							&	\nodata					&	$	4.06 	_{-	0.40 	}^{+	0.43 	}$	&	$	19.75 	_{-	2.98 	}^{+	3.69 	}$	&	$	197.99 	/	198 	$	\\
20		& Nov 14 00:30:46.836	&	0.081	&$	0.24 	\pm	0.04 	$&	\nodata							&	\nodata							&	\nodata					&	$	9.00 	_{-	1.14 	}^{+	1.28 	}$	&	\nodata							&	$	225.80 	/	200 	$	\\
21	$^{T}$	& Nov 14 19:50:42.295	&	0.024	&$	0.55 	\pm	0.04 	$&	\nodata							&	\nodata							&	\nodata					&	$	4.19 	_{-	0.54 	}^{+	0.56 	}$	&	$	13.56 	_{-	1.80 	}^{+	2.28 	}$	&	$	213.79 	/	199 	$	\\
22	$^{T}$	& Nov 15 20:48:41.297	&	0.037	&$	0.79 	\pm	0.06 	$&	\nodata							&	\nodata							&	\nodata					&	$	3.76 	_{-	0.38 	}^{+	0.42 	}$	&	$	14.04 	_{-	1.71 	}^{+	2.07 	}$	&	$	128.83 	/	130 	$	\\
\cline{1-10}
\multicolumn{10}{c}{2020} \\
\cline{1-10}
23	$^{T}$	& Apr 10 09:43:54.273	&	0.171	&$	27.46 	\pm	0.41 	$&	$	32.76 	_{-	0.38 	}^{+	0.38 	}$	&	$	0.85 	_{-	0.11 	}^{+	0.11 	}$	&	$	127.16 	/	65 	$	&	$	6.81 	_{-	0.92 	}^{+	0.70 	}$	&	$	11.05 	_{-	1.10 	}^{+	2.15 	}$	&	$	126.27 	/	64 	$	\\
24	$^{T}$	& Apr 27 18:26:20.138	&	0.216	&$	2.05 	\pm	0.10 	$&	$	26.25 	_{-	1.61 	}^{+	1.55 	}$	&	$	-0.36 	_{-	0.34 	}^{+	0.37 	}$	&	$	154.10 	/	134 	$	&	$	4.33 	_{-	0.74 	}^{+	0.78 	}$	&	$	10.13 	_{-	1.27 	}^{+	2.05 	}$	&	$	153.37 	/	133 	$	\\
25		& Apr 27 18:31:05.770	&	0.244	&$	0.76 	\pm	0.07 	$&	\nodata							&	\nodata							&	\nodata					&	$	3.47 	_{-	1.07 	}^{+	0.85 	}$	&	$	10.99 	_{-	2.48 	}^{+	2.85 	}$	&	$	185.00 	/	200 	$	\\
26		& Apr 27 18:31:25.234	&	0.166	&$	1.89 	\pm	0.08 	$&	$	23.83 	_{-	1.95 	}^{+	1.70 	}$	&	$	-0.75 	_{-	0.28 	}^{+	0.30 	}$	&	$	240.78 	/	201 	$	&	$	4.15 	_{-	0.56 	}^{+	0.53 	}$	&	$	10.49 	_{-	1.17 	}^{+	1.50 	}$	&	$	240.93 	/	200 	$	\\
27		& Apr 27 18:33:53.116	&	0.071	&$	0.51 	\pm	0.04 	$&	\nodata							&	$	-2.39 	_{-	0.14 	}^{+	0.13 	}$	&	$	303.72 	/	269 	$	&	\nodata							&	\nodata							&	\nodata					\\
28		& Apr 27 18:34:05.700	&	0.422	&$	20.11 	\pm	0.19 	$&	$	27.29 	_{-	0.27 	}^{+	0.27 	}$	&	$	0.23 	_{-	0.08 	}^{+	0.08 	}$	&	$	392.00 	/	268 	$	&	$	5.53 	_{-	0.20 	}^{+	0.19 	}$	&	$	10.90 	_{-	0.51 	}^{+	0.57 	}$	&	$	367.22 	/	267 	$	\\
29		& Apr 27 18:34:46.047	&	0.226	&$	3.83 	\pm	0.11 	$&	\nodata							&	\nodata							&	\nodata					&	$	4.45 	_{-	0.19 	}^{+	0.19 	}$	&	$	14.30 	_{-	1.10 	}^{+	1.23 	}$	&	$	211.64 	/	198 	$	\\
30		& Apr 27 18:34:47.296	&	0.534	&$	1.05 	\pm	0.10 	$&	\nodata							&	\nodata							&	\nodata					&	$	3.96 	_{-	0.58 	}^{+	0.58 	}$	&	$	13.99 	_{-	2.65 	}^{+	3.34 	}$	&	$	203.53 	/	198 	$	\\
31		& Apr 27 18:35:05.320	&	0.103	&$	5.12 	\pm	0.11 	$&	$	27.36 	_{-	0.73 	}^{+	0.71 	}$	&	$	-0.24 	_{-	0.15 	}^{+	0.16 	}$	&	$	220.85 	/	199 	$	&	$	4.99 	_{-	0.35 	}^{+	0.34 	}$	&	$	11.06 	_{-	0.71 	}^{+	0.84 	}$	&	$	217.28 	/	198 	$	\\
32		& Apr 27 18:35:46.623	&	0.061	&$	1.32 	\pm	0.06 	$&	$	27.56 	_{-	1.12 	}^{+	1.12 	}$	&	$	0.58 	_{-	0.37 	}^{+	0.40 	}$	&	$	188.67 	/	199 	$	&	\nodata							&	\nodata							&	\nodata					\\
33		& Apr 27 18:35:57.633	&	0.025	&$	0.45 	\pm	0.04 	$&	$	19.46 	_{-	2.22 	}^{+	2.59 	}$	&	\nodata							&	$	194.59 	/	200 	$	&	\nodata							&	\nodata							&	\nodata					\\
34		& Apr 27 18:36:45.376	&	0.014	&$	0.28 	\pm	0.03 	$&	$	19.26 	_{-	2.41 	}^{+	2.86 	}$	&	\nodata							&	$	247.99 	/	267 	$	&	\nodata							&	\nodata							&	\nodata					\\
35	$^{T}$	& Apr 27 18:36:46.007	&	0.346	&$	19.42 	\pm	0.20 	$&	$	29.26 	_{-	0.35 	}^{+	0.34 	}$	&	$	-0.12 	_{-	0.07 	}^{+	0.08 	}$	&	$	364.99 	/	266 	$	&	$	4.97 	_{-	0.24 	}^{+	0.23 	}$	&	$	10.76 	_{-	0.35 	}^{+	0.37 	}$	&	$	374.53 	/	265 	$	\\
36		& Apr 27 18:38:20.206	&	0.105	&$	0.93 	\pm	0.06 	$&	\nodata							&	\nodata							&	\nodata					&	$	4.03 	_{-	0.31 	}^{+	0.33 	}$	&	$	17.67 	_{-	2.45 	}^{+	2.89 	}$	&	$	423.22 	/	332 	$	\\
37		& Apr 27 18:38:53.689	&	0.250	&$	1.81 	\pm	0.08 	$&	\nodata							&	\nodata							&	\nodata					&	$	4.66 	_{-	0.35 	}^{+	0.33 	}$	&	$	17.14 	_{-	2.31 	}^{+	2.66 	}$	&	$	351.02 	/	332 	$	\\
38		& Apr 27 18:39:09.331	&	0.035	&$	0.19 	\pm	0.03 	$&	\nodata							&	$	-2.11 	_{-	0.19 	}^{+	0.19 	}$	&	$	320.11 	/	334 	$	&	\nodata							&	\nodata							&	\nodata					\\
39		& Apr 27 18:40:15.043	&	0.456	&$	1.15 	\pm	0.09 	$&	$	23.65 	_{-	3.06 	}^{+	3.49 	}$	&	\nodata							&	$	305.48 	/	266 	$	&	\nodata							&	\nodata							&	\nodata					\\
40		& Apr 27 18:40:32.031	&	1.353	&$	6.16 	\pm	0.18 	$&	$	24.74 	_{-	1.74 	}^{+	1.51 	}$	&	$	-1.07 	_{-	0.18 	}^{+	0.19 	}$	&	$	375.27 	/	265 	$	&	$	4.19 	_{-	0.28 	}^{+	0.28 	}$	&	$	11.71 	_{-	0.80 	}^{+	0.93 	}$	&	$	371.44 	/	264 	$	\\
41		& Apr 27 18:42:40.816	&	0.031	&$	0.35 	\pm	0.03 	$&	$	14.01 	_{-	2.05 	}^{+	2.52 	}$	&	\nodata							&	$	173.11 	/	199 	$	&	\nodata							&	\nodata							&	\nodata					\\
42		& Apr 27 18:42:50.652	&	0.316	&$	8.17 	\pm	0.15 	$&	$	23.44 	_{-	0.92 	}^{+	0.85 	}$	&	$	-0.81 	_{-	0.13 	}^{+	0.14 	}$	&	$	274.17 	/	198 	$	&	$	4.86 	_{-	0.19 	}^{+	0.18 	}$	&	$	12.86 	_{-	0.79 	}^{+	0.87 	}$	&	$	256.98 	/	197 	$	\\
43		& Apr 27 18:44:08.209	&	0.077	&$	0.34 	\pm	0.06 	$&	\nodata							&	\nodata							&	\nodata					&	$	2.98 	_{-	0.62 	}^{+	0.78 	}$	&	$	14.69 	_{-	3.07 	}^{+	4.01 	}$	&	$	139.07 	/	130 	$	\\
44		& Apr 27 18:46:08.767	&	0.206	&$	3.86 	\pm	0.13 	$&	$	25.91 	_{-	1.20 	}^{+	1.13 	}$	&	$	-0.40 	_{-	0.25 	}^{+	0.26 	}$	&	$	140.86 	/	131 	$	&	$	5.06 	_{-	0.59 	}^{+	0.48 	}$	&	$	11.75 	_{-	1.61 	}^{+	1.93 	}$	&	$	140.74 	/	130 	$	\\
45		& Apr 27 18:46:39.414	&	1.651	&$	9.70 	\pm	0.26 	$&	$	25.08 	_{-	0.90 	}^{+	0.86 	}$	&	$	-0.40 	_{-	0.21 	}^{+	0.22 	}$	&	$	185.19 	/	131 	$	&	$	4.98 	_{-	1.13 	}^{+	0.40 	}$	&	$	11.83 	_{-	2.92 	}^{+	2.10 	}$	&	$	185.35 	/	130 	$	\\
46	$^{T}$	& Apr 27 18:47:05.754	&	0.155	&$	11.78 	\pm	0.20 	$&	$	31.86 	_{-	0.47 	}^{+	0.47 	}$	&	$	0.56 	_{-	0.13 	}^{+	0.14 	}$	&	$	165.91 	/	131 	$	&	$	5.21 	_{-	0.64 	}^{+	0.67 	}$	&	$	9.92 	_{-	0.45 	}^{+	0.64 	}$	&	$	165.95 	/	130 	$	\\
47		& Apr 27 18:48:38.675	&	0.243	&$	1.69 	\pm	0.10 	$&	\nodata							&	\nodata							&	\nodata					&	$	4.10 	_{-	0.68 	}^{+	0.60 	}$	&	$	11.84 	_{-	1.95 	}^{+	2.48 	}$	&	$	129.68 	/	130 	$	\\
48		& Apr 27 18:49:28.034	&	0.368	&$	6.83 	\pm	0.17 	$&	$	23.06 	_{-	1.16 	}^{+	1.05 	}$	&	$	-0.75 	_{-	0.18 	}^{+	0.19 	}$	&	$	189.01 	/	131 	$	&	$	4.58 	_{-	0.27 	}^{+	0.26 	}$	&	$	11.66 	_{-	0.93 	}^{+	1.06 	}$	&	$	181.99 	/	130 	$	\\
49		& Apr 27 18:50:28.665	&	0.025	&$	0.27 	\pm	0.03 	$&	\nodata							&	\nodata							&	\nodata					&	$	6.14 	_{-	0.60 	}^{+	0.68 	}$	&	\nodata							&	$	118.79 	/	132 	$	\\
50		& Apr 27 18:50:49.460	&	0.035	&$	0.26 	\pm	0.04 	$&	\nodata							&	\nodata							&	\nodata					&	$	7.38 	_{-	0.78 	}^{+	0.85 	}$	&	\nodata							&	$	107.17 	/	132 	$	\\
51		& Apr 27 18:55:44.155	&	0.072	&$	2.65 	\pm	0.12 	$&	$	28.91 	_{-	1.95 	}^{+	1.79 	}$	&	$	-0.62 	_{-	0.27 	}^{+	0.29 	}$	&	$	132.47 	/	131 	$	&	$	5.26 	_{-	0.50 	}^{+	0.49 	}$	&	$	13.46 	_{-	1.48 	}^{+	1.82 	}$	&	$	129.30 	/	130 	$	\\
52		& Apr 27 18:57:35.574	&	0.096	&$	0.78	\pm	0.08	$&	$	26.42	_{-	3.87	}^{+	4.62	}$	&	\nodata							&	$	148.87	/	132 	$	&	\nodata							&	\nodata							&	\nodata					\\
53		& Apr 27 18:58:45.533	&	0.193	&$	0.78	\pm	0.11	$&	\nodata							&	$	-2.397	_{-	0.222	}^{+	0.193	}$	&	$	166	/	132 	$	&	\nodata							&	\nodata							&	\nodata					\\
54		& Apr 27 19:36:05.104	&	0.028	&$	0.40 	\pm	0.04 	$&	\nodata							&	\nodata							&	\nodata					&	$	4.51 	_{-	0.49 	}^{+	0.54 	}$	&	$	20.09 	_{-	4.21 	}^{+	6.04 	}$	&	$	207.13 	/	199 	$	\\
55	$^{T}$	& Apr 27 19:37:39.328	&	0.724	&$	3.89 	\pm	0.16 	$&	\nodata							&	\nodata							&	\nodata					&	$	4.29 	_{-	0.24 	}^{+	0.24 	}$	&	$	16.26 	_{-	1.63 	}^{+	1.82 	}$	&	$	238.87 	/	199 	$	\\
56		& Apr 27 19:43:44.537	&	0.436	&$	38.42 	\pm	0.24 	$&	$	32.02 	_{-	0.17 	}^{+	0.17 	}$	&	$	0.78 	_{-	0.05 	}^{+	0.05 	}$	&	$	591.99 	/	337 	$	&	$	6.91 	_{-	0.27 	}^{+	0.23 	}$	&	$	11.47 	_{-	0.58 	}^{+	0.70 	}$	&	$	583.95 	/	336 	$	\\
57		& Apr 27 19:45:00.478	&	0.101	&$	0.51 	\pm	0.05 	$&	\nodata							&	\nodata							&	\nodata					&	$	1.94 	_{-	0.38 	}^{+	0.46 	}$	&	$	9.69 	_{-	1.02 	}^{+	1.22 	}$	&	$	379.07 	/	336 	$	\\
58		& Apr 27 19:55:32.325	&	0.041	&$	0.23 	\pm	0.03 	$&	\nodata							&	\nodata							&	\nodata					&	$	5.35 	_{-	0.47 	}^{+	0.52 	}$	&	\nodata							&	$	179.42 	/	203 	$	\\
59	$^{T}$	& Apr 27 20:01:45.681	&	0.483	&$	5.08 	\pm	0.15 	$&	$	26.24 	_{-	1.11 	}^{+	1.06 	}$	&	$	-0.50 	_{-	0.21 	}^{+	0.22 	}$	&	$	155.70 	/	134 	$	&	$	4.55 	_{-	0.41 	}^{+	0.39 	}$	&	$	10.92 	_{-	0.94 	}^{+	1.11 	}$	&	$	154.88 	/	133 	$	\\
60		& Apr 27 20:07:20.319	&	0.382	&$	0.79 	\pm	0.10 	$&	$	56.43 	_{-	10.20 	}^{+	14.00 	}$	&	\nodata							&	$	219.88 	/	202 	$	&	\nodata							&	\nodata							&	\nodata					\\
61	$^{T}$	& Apr 27 20:13:38.263	&	0.055	&$	0.52 	\pm	0.04 	$&	\nodata							&	\nodata							&	\nodata					&	$	4.29 	_{-	1.13 	}^{+	1.16 	}$	&	$	10.83 	_{-	1.59 	}^{+	2.52 	}$	&	$	328.39 	/	332 	$	\\
62		& Apr 27 20:14:51.396	&	0.051	&$	1.38 	\pm	0.06 	$&	$	27.73 	_{-	1.75 	}^{+	1.61 	}$	&	$	-0.54 	_{-	0.28 	}^{+	0.29 	}$	&	$	267.85 	/	265 	$	&	$	4.44 	_{-	0.54 	}^{+	0.57 	}$	&	$	11.07 	_{-	0.96 	}^{+	1.24 	}$	&	$	266.26 	/	264 	$	\\
63		& Apr 27 20:15:20.583	&	1.282	&$	189.77 	\pm	0.67 	$&	$	34.25 	_{-	0.10 	}^{+	0.10 	}$	&	$	0.56 	_{-	0.03 	}^{+	0.03 	}$	&	$	639.98 	/	247 	$	&	$	6.61 	_{-	0.15 	}^{+	0.17 	}$	&	$	11.77 	_{-	0.21 	}^{+	0.25 	}$	&	$	630.40 	/	246 	$	\\
64		& Apr 27 20:16:15.285	&	0.030	&$	0.32 	\pm	0.03 	$&	$	21.49 	_{-	2.93 	}^{+	3.38 	}$	&	\nodata							&	$	271.06 	/	266 	$	&	\nodata							&	\nodata							&	\nodata					\\
65		& Apr 27 20:17:09.139	&	0.110	&$	0.71 	\pm	0.06 	$&	\nodata							&	\nodata							&	\nodata					&	$	3.94 	_{-	0.53 	}^{+	0.49 	}$	&	$	12.32 	_{-	2.05 	}^{+	2.51 	}$	&	$	226.45 	/	197 	$	\\
66		& Apr 27 20:17:27.317	&	0.064	&$	0.58 	\pm	0.05 	$&	\nodata							&	\nodata							&	\nodata					&	$	5.09 	_{-	0.93 	}^{+	0.83 	}$	&	$	18.29 	_{-	5.25 	}^{+	6.82 	}$	&	$	171.55 	/	197 	$	\\
67		& Apr 27 20:17:50.343	&	1.422	&$	2.20 	\pm	0.18 	$&	$	36.27 	_{-	5.03 	}^{+	6.13 	}$	&	\nodata							&	$	275.72 	/	199 	$	&	\nodata							&	\nodata							&	\nodata					\\
68		& Apr 27 20:17:58.442	&	0.078	&$	1.15 	\pm	0.06 	$&	\nodata							&	\nodata							&	\nodata					&	$	4.42 	_{-	0.76 	}^{+	0.66 	}$	&	$	11.19 	_{-	1.70 	}^{+	2.25 	}$	&	$	219.72 	/	197 	$	\\
69		& Apr 27 20:18:09.130	&	0.836	&$	0.36 	\pm	0.09 	$&	\nodata							&	\nodata							&	\nodata					&	$	7.84 	_{-	1.44 	}^{+	1.83 	}$	&	\nodata							&	$	214.22 	/	199 	$	\\
70		& Apr 27 20:19:23.068	&	0.030	&$	0.31 	\pm	0.04 	$&	\nodata							&	$	-2.71 	_{-	0.25 	}^{+	0.22 	}$	&	$	121.56 	/	132 	$	&	\nodata							&	\nodata							&	\nodata					\\
71		& Apr 27 20:19:47.631	&	0.849	&$	1.16 	\pm	0.17 	$&	\nodata							&	$	-1.92 	_{-	0.18 	}^{+	0.17 	}$	&	$	131.42 	/	132 	$	&	\nodata							&	\nodata							&	\nodata					\\
72		& Apr 27 20:19:49.430	&	0.232	&$	4.09 	\pm	0.14 	$&	$	23.83 	_{-	1.83 	}^{+	1.55 	}$	&	$	-0.93 	_{-	0.24 	}^{+	0.25 	}$	&	$	151.93 	/	131 	$	&	$	4.97 	_{-	0.43 	}^{+	0.36 	}$	&	$	14.06 	_{-	2.10 	}^{+	2.35 	}$	&	$	147.39 	/	130 	$	\\
73		& Apr 27 20:20:44.640	&	0.335	&$	0.74	\pm	0.09	$&	$	23.79	_{-	4.58	}^{+	5.85	}$	&	\nodata							&	$	121.99	/	132 	$	&	\nodata							&	\nodata							&	\nodata					\\
74		& Apr 27 20:21:51.841	&	0.578	&$	1.63 	\pm	0.14 	$&	\nodata							&	\nodata							&	\nodata					&	$	3.91 	_{-	0.43 	}^{+	0.49 	}$	&	$	16.33 	_{-	2.10 	}^{+	2.44 	}$	&	$	179.22 	/	130 	$	\\
75		& Apr 27 20:21:55.136	&	0.549	&$	9.34 	\pm	0.20 	$&	$	23.67 	_{-	0.80 	}^{+	0.75 	}$	&	$	-0.48 	_{-	0.17 	}^{+	0.18 	}$	&	$	166.74 	/	131 	$	&	$	4.81 	_{-	0.39 	}^{+	0.32 	}$	&	$	11.46 	_{-	1.32 	}^{+	1.52 	}$	&	$	162.53 	/	130 	$	\\
76		& Apr 27 20:25:53.415	&	0.400	&$	1.55 	\pm	0.14 	$&	\nodata							&	\nodata							&	\nodata					&	$	4.34 	_{-	0.49 	}^{+	0.53 	}$	&	$	16.82 	_{-	2.72 	}^{+	3.52 	}$	&	$	148.97 	/	130 	$	\\
77	$^{T}$	& Apr 27 21:14:45.605	&	0.265	&$	10.15 	\pm	0.15 	$&	$	30.93 	_{-	0.48 	}^{+	0.47 	}$	&	$	0.01 	_{-	0.10 	}^{+	0.10 	}$	&	$	366.35 	/	268 	$	&	$	6.12 	_{-	0.31 	}^{+	0.30 	}$	&	$	12.78 	_{-	0.78 	}^{+	1.01 	}$	&	$	363.69 	/	267 	$	\\
78		& Apr 27 21:15:36.398	&	0.383	&$	6.47 	\pm	0.13 	$&	$	27.39 	_{-	0.81 	}^{+	0.78 	}$	&	$	-0.53 	_{-	0.14 	}^{+	0.15 	}$	&	$	309.75 	/	268 	$	&	$	4.84 	_{-	0.28 	}^{+	0.27 	}$	&	$	11.70 	_{-	0.70 	}^{+	0.79 	}$	&	$	305.47 	/	267 	$	\\
79		& Apr 27 21:20:55.561	&	0.089	&$	0.96 	\pm	0.05 	$&	\nodata							&	\nodata							&	\nodata					&	$	4.16 	_{-	0.48 	}^{+	0.42 	}$	&	$	10.23 	_{-	1.56 	}^{+	1.93 	}$	&	$	409.23 	/	336 	$	\\
80		& Apr 27 21:20:58.670	&	0.187	&$	0.77 	\pm	0.06 	$&	\nodata							&	\nodata							&	\nodata					&	$	4.34 	_{-	0.41 	}^{+	0.44 	}$	&	$	16.00 	_{-	2.52 	}^{+	3.13 	}$	&	$	396.53 	/	336 	$	\\
81		& Apr 27 21:24:05.936	&	0.050	&$	0.42 	\pm	0.04 	$&	\nodata							&	$	-2.17 	_{-	0.15 	}^{+	0.14 	}$	&	$	221.69 	/	203 	$	&	\nodata							&	\nodata							&	\nodata					\\
82		& Apr 27 21:25:01.037	&	0.060	&$	0.39 	\pm	0.04 	$&	$	24.03 	_{-	3.19 	}^{+	3.83 	}$	&	\nodata							&	$	190.57 	/	203 	$	&	\nodata							&	\nodata							&	\nodata					\\
83		& Apr 27 21:27:25.367	&	0.246	&$	0.47 	\pm	0.07 	$&	\nodata							&	$	-2.15 	_{-	0.19 	}^{+	0.17 	}$	&	$	242.45 	/	203 	$	&	\nodata							&	\nodata							&	\nodata					\\
84	$^{T}$	& Apr 27 21:43:06.346	&	0.163	&$	1.66 	\pm	0.08 	$&	$	17.05 	_{-	1.35 	}^{+	1.50 	}$	&	\nodata							&	$	152.64 	/	134 	$	&	\nodata							&	\nodata							&	\nodata					\\
85		& Apr 27 21:48:44.062	&	0.283	&$	6.59 	\pm	0.12 	$&	$	23.76 	_{-	0.70 	}^{+	0.66 	}$	&	$	-0.36 	_{-	0.15 	}^{+	0.15 	}$	&	$	304.76 	/	266 	$	&	$	4.31 	_{-	0.31 	}^{+	0.29 	}$	&	$	9.57 	_{-	0.53 	}^{+	0.61 	}$	&	$	304.16 	/	265 	$	\\
86		& Apr 27 21:57:03.989	&	0.029	&$	0.23 	\pm	0.04 	$&	\nodata							&	$	-2.70 	_{-	0.32 	}^{+	0.27 	}$	&	$	147.94 	/	132 	$	&	\nodata							&	\nodata							&	\nodata					\\
87	$^{T}$	& Apr 27 21:59:22.528	&	0.239	&$	14.88 	\pm	0.23 	$&	$	32.32 	_{-	0.48 	}^{+	0.48 	}$	&	$	0.22 	_{-	0.11 	}^{+	0.12 	}$	&	$	143.43 	/	131 	$	&	$	5.21 	_{-	0.41 	}^{+	0.41 	}$	&	$	10.81 	_{-	0.44 	}^{+	0.53 	}$	&	$	142.45 	/	130 	$	\\
88		& Apr 27 22:47:05.343	&	0.017	&$	0.20 	\pm	0.03 	$&	$	28.94 	_{-	6.51 	}^{+	8.85 	}$	&	\nodata							&	$	185.79 	/	201 	$	&	\nodata							&	\nodata							&	\nodata					\\
89	$^{T}$	& Apr 27 22:55:19.911	&	0.266	&$	1.97 	\pm	0.10 	$&	\nodata							&	\nodata							&	\nodata					&	$	5.17 	_{-	0.47 	}^{+	0.47 	}$	&	$	17.42 	_{-	2.56 	}^{+	3.34 	}$	&	$	228.90 	/	201 	$	\\
90		& Apr 27 23:02:53.488	&	0.261	&$	7.10 	\pm	0.13 	$&	$	22.37 	_{-	0.72 	}^{+	0.67 	}$	&	$	-0.52 	_{-	0.14 	}^{+	0.15 	}$	&	$	252.34 	/	202 	$	&	$	4.26 	_{-	0.25 	}^{+	0.24 	}$	&	$	9.80 	_{-	0.55 	}^{+	0.63 	}$	&	$	249.33 	/	201 	$	\\
91	$^{T}$	& Apr 27 23:06:06.135	&	0.166	&$	2.02 	\pm	0.08 	$&	$	24.68 	_{-	1.71 	}^{+	1.53 	}$	&	$	-0.67 	_{-	0.26 	}^{+	0.28 	}$	&	$	185.58 	/	202 	$	&	$	4.50 	_{-	0.37 	}^{+	0.38 	}$	&	$	11.50 	_{-	1.09 	}^{+	1.30 	}$	&	$	180.17 	/	201 	$	\\
92	$^{T}$	& Apr 27 23:25:04.349	&	0.502	&$	1.75 	\pm	0.11 	$&	\nodata							&	$	-2.22 	_{-	0.09 	}^{+	0.08 	}$	&	$	352.76 	/	266 	$	&	\nodata							&	\nodata							&	\nodata					\\
93		& Apr 27 23:27:46.293	&	0.068	&$	3.23 	\pm	0.09 	$&	$	29.05 	_{-	0.77 	}^{+	0.76 	}$	&	$	0.45 	_{-	0.23 	}^{+	0.25 	}$	&	$	207.07 	/	198 	$	&	$	6.26 	_{-	0.67 	}^{+	0.49 	}$	&	$	12.27 	_{-	1.92 	}^{+	2.74 	}$	&	$	203.33 	/	197 	$	\\
94	$^{T}$	& Apr 27 23:42:41.143	&	0.053	&$	1.93 	\pm	0.10 	$&	$	28.75 	_{-	1.77 	}^{+	1.67 	}$	&	$	-0.11 	_{-	0.37 	}^{+	0.40 	}$	&	$	153.33 	/	131 	$	&	$	3.99 	_{-	0.63 	}^{+	0.57 	}$	&	$	9.83 	_{-	0.76 	}^{+	0.82 	}$	&	$	150.21 	/	130 	$	\\
95		& Apr 27 23:44:31.818	&	0.322	&$	14.49 	\pm	0.30 	$&	$	38.69 	_{-	0.63 	}^{+	0.64 	}$	&	$	0.84 	_{-	0.16 	}^{+	0.17 	}$	&	$	189.12 	/	131 	$	&	$	8.56 	_{-	0.30 	}^{+	0.30 	}$	&	$	18.98 	_{-	1.92 	}^{+	2.47 	}$	&	$	158.25 	/	130 	$	\\
96	$^{T}$	& Apr 28 00:19:44.173	&	0.151	&$	1.11 	\pm	0.06 	$&	$	20.62 	_{-	1.94 	}^{+	2.17 	}$	&	\nodata							&	$	251.63 	/	201 	$	&	\nodata							&	\nodata							&	\nodata					\\
97		& Apr 28 00:23:04.763	&	0.113	&$	0.69 	\pm	0.06 	$&	\nodata							&	\nodata							&	\nodata					&	$	3.01 	_{-	0.46 	}^{+	0.54 	}$	&	$	10.19 	_{-	1.11 	}^{+	1.39 	}$	&	$	191.30 	/	199 	$	\\
98		& Apr 28 00:24:30.311	&	0.236	&$	34.49 	\pm	0.26 	$&	$	32.72 	_{-	0.21 	}^{+	0.21 	}$	&	$	0.66 	_{-	0.06 	}^{+	0.06 	}$	&	$	409.07 	/	268 	$	&	$	6.46 	_{-	0.37 	}^{+	0.34 	}$	&	$	11.02 	_{-	0.48 	}^{+	0.59 	}$	&	$	410.88 	/	267 	$	\\
99		& Apr 28 00:25:43.946	&	0.042	&$	0.25 	\pm	0.03 	$&	\nodata							&	\nodata							&	\nodata					&	$	2.68 	_{-	0.60 	}^{+	0.58 	}$	&	$	11.57 	_{-	3.02 	}^{+	3.65 	}$	&	$	267.66 	/	267 	$	\\
100		& Apr 28 00:37:36.160	&	0.115	&$	0.63 	\pm	0.05 	$&	$	25.87 	_{-	3.32 	}^{+	3.89 	}$	&	\nodata							&	$	214.84 	/	203 	$	&	\nodata							&	\nodata							&	\nodata					\\
101	$^{T}$	& Apr 28 00:39:39.565	&	0.659	&$	2.29 	\pm	0.11 	$&	\nodata							&	\nodata							&	\nodata					&	$	3.67 	_{-	0.50 	}^{+	0.49 	}$	&	$	10.14 	_{-	1.38 	}^{+	1.92 	}$	&	$	209.97 	/	201 	$	\\
102		& Apr 28 00:40:33.077	&	0.689	&$	4.28 	\pm	0.13 	$&	\nodata							&	\nodata							&	\nodata					&	$	3.88 	_{-	0.22 	}^{+	0.22 	}$	&	$	11.30 	_{-	0.97 	}^{+	1.09 	}$	&	$	268.16 	/	201 	$	\\
103		& Apr 28 00:41:32.148	&	0.437	&$	13.76 	\pm	0.19 	$&	$	27.22 	_{-	0.51 	}^{+	0.50 	}$	&	$	-0.44 	_{-	0.10 	}^{+	0.10 	}$	&	$	271.86 	/	202 	$	&	$	5.10 	_{-	0.29 	}^{+	0.24 	}$	&	$	11.71 	_{-	0.70 	}^{+	0.71 	}$	&	$	286.28 	/	201 	$	\\
104		& Apr 28 00:43:24.784	&	0.846	&$	6.52 	\pm	0.15 	$&	$	22.58 	_{-	0.60 	}^{+	0.58 	}$	&	$	0.26 	_{-	0.23 	}^{+	0.23 	}$	&	$	278.24 	/	202 	$	&	$	4.66 	_{-	0.38 	}^{+	0.34 	}$	&	$	9.18 	_{-	0.97 	}^{+	1.37 	}$	&	$	271.59 	/	201 	$	\\
105		& Apr 28 00:44:08.210	&	1.275	&$	70.82 	\pm	0.45 	$&	$	38.45 	_{-	0.21 	}^{+	0.21 	}$	&	$	0.29 	_{-	0.04 	}^{+	0.04 	}$	&	$	569.03 	/	202 	$	&	$	6.35 	_{-	0.24 	}^{+	0.22 	}$	&	$	12.80 	_{-	0.27 	}^{+	0.28 	}$	&	$	597.48 	/	201 	$	\\
106		& Apr 28 00:45:31.098	&	0.107	&$	0.86 	\pm	0.06 	$&	$	21.21 	_{-	2.05 	}^{+	2.33 	}$	&	\nodata							&	$	213.69 	/	203 	$	&	\nodata							&	\nodata							&	\nodata					\\
107		& Apr 28 00:46:00.034	&	0.798	&$	6.05 	\pm	0.18 	$&	$	23.09 	_{-	1.96 	}^{+	1.68 	}$	&	$	-1.19 	_{-	0.16 	}^{+	0.17 	}$	&	$	207.89 	/	134 	$	&	$	5.17 	_{-	0.22 	}^{+	0.22 	}$	&	$	17.32 	_{-	1.54 	}^{+	1.78 	}$	&	$	188.72 	/	133 	$	\\
108		& Apr 28 00:46:06.394	&	0.176	&$	0.87 	\pm	0.07 	$&	$	22.12 	_{-	2.74 	}^{+	3.19 	}$	&	\nodata							&	$	156.89 	/	135 	$	&	\nodata							&	\nodata							&	\nodata					\\
109		& Apr 28 00:46:20.179	&	0.851	&$	46.93 	\pm	0.43 	$&	$	38.72 	_{-	0.31 	}^{+	0.31 	}$	&	$	0.27 	_{-	0.06 	}^{+	0.06 	}$	&	$	240.75 	/	134 	$	&	$	6.85 	_{-	0.29 	}^{+	0.31 	}$	&	$	13.80 	_{-	0.49 	}^{+	0.62 	}$	&	$	242.40 	/	133 	$	\\
110		& Apr 28 00:46:23.528	&	0.843	&$	2.71 	\pm	0.14 	$&	$	17.90 	_{-	1.20 	}^{+	1.00 	}$	&	$	0.35 	_{-	0.59 	}^{+	0.72 	}$	&	$	138.48 	/	134 	$	&	\nodata							&	\nodata							&	\nodata					\\
111		& Apr 28 00:46:43.072	&	0.503	&$	3.82 	\pm	0.14 	$&	$	23.23 	_{-	1.09 	}^{+	1.03 	}$	&	$	-0.19 	_{-	0.30 	}^{+	0.32 	}$	&	$	146.59 	/	134 	$	&	$	4.32 	_{-	0.55 	}^{+	0.52 	}$	&	$	9.35 	_{-	1.09 	}^{+	1.63 	}$	&	$	144.69 	/	133 	$	\\
112		& Apr 28 00:47:24.957	&	0.239	&$	0.84 	\pm	0.08 	$&	$	21.07 	_{-	2.84 	}^{+	3.34 	}$	&	\nodata							&	$	142.74 	/	135 	$	&	\nodata							&	\nodata							&	\nodata					\\
113		& Apr 28 00:47:57.536	&	0.260	&$	1.50 	\pm	0.09 	$&	$	21.31 	_{-	1.38 	}^{+	1.33 	}$	&	$	0.29 	_{-	0.56 	}^{+	0.61 	}$	&	$	141.30 	/	134 	$	&	\nodata							&	\nodata							&	\nodata					\\
114		& Apr 28 00:48:44.833	&	0.443	&$	1.88 	\pm	0.11 	$&	\nodata							&	\nodata							&	\nodata					&	$	4.58 	_{-	0.27 	}^{+	0.26 	}$	&	$	16.25 	_{-	3.11 	}^{+	3.91 	}$	&	$	216.35 	/	200 	$	\\
115		& Apr 28 00:48:49.098	&	0.597	&$	8.17 	\pm	0.17 	$&	$	31.17 	_{-	0.94 	}^{+	0.93 	}$	&	$	-0.75 	_{-	0.12 	}^{+	0.12 	}$	&	$	265.81 	/	201 	$	&	$	5.46 	_{-	0.25 	}^{+	0.25 	}$	&	$	14.91 	_{-	0.91 	}^{+	1.04 	}$	&	$	260.72 	/	200 	$	\\
116		& Apr 28 00:49:00.270	&	2.607	&$	2.85 	\pm	0.21 	$&	\nodata							&	\nodata							&	\nodata					&	$	3.63 	_{-	0.39 	}^{+	0.40 	}$	&	$	11.76 	_{-	1.59 	}^{+	1.93 	}$	&	$	356.21 	/	200 	$	\\
117		& Apr 28 00:49:06.479	&	0.027	&$	0.21 	\pm	0.03 	$&	$	15.12 	_{-	2.58 	}^{+	3.23 	}$	&	\nodata							&	$	150.25 	/	202 	$	&	\nodata							&	\nodata							&	\nodata					\\
118		& Apr 28 00:49:16.609	&	0.313	&$	3.94 	\pm	0.11 	$&	$	20.64 	_{-	0.99 	}^{+	0.88 	}$	&	$	-0.40 	_{-	0.26 	}^{+	0.29 	}$	&	$	218.99 	/	201 	$	&	$	4.18 	_{-	0.21 	}^{+	0.19 	}$	&	$	14.39 	_{-	1.92 	}^{+	2.22 	}$	&	$	195.58 	/	200 	$	\\
119		& Apr 28 00:49:22.392	&	0.091	&$	0.64 	\pm	0.04 	$&	$	14.28 	_{-	1.60 	}^{+	1.88 	}$	&	\nodata							&	$	241.30 	/	202 	$	&	\nodata							&	\nodata							&	\nodata					\\
120		& Apr 28 00:49:27.008	&	0.347	&$	0.64 	\pm	0.07 	$&	$	21.46 	_{-	3.83 	}^{+	4.71 	}$	&	\nodata							&	$	270.72 	/	202 	$	&	\nodata							&	\nodata							&	\nodata					\\
121	$^{T}$	& Apr 28 00:49:45.895	&	1.164	&$	2.88 	\pm	0.16 	$&	\nodata							&	\nodata							&	\nodata					&	$	3.97 	_{-	0.25 	}^{+	0.25 	}$	&	$	16.98 	_{-	2.43 	}^{+	2.83 	}$	&	$	259.29 	/	197 	$	\\
122		& Apr 28 00:50:01.012	&	0.499	&$	3.64 	\pm	0.11 	$&	$	21.34 	_{-	0.82 	}^{+	0.76 	}$	&	$	0.08 	_{-	0.30 	}^{+	0.31 	}$	&	$	173.68 	/	201 	$	&	$	4.65 	_{-	0.46 	}^{+	0.35 	}$	&	$	9.82 	_{-	1.60 	}^{+	2.14 	}$	&	$	170.29 	/	200 	$	\\
123		& Apr 28 00:50:21.993	&	0.021	&$	0.18 	\pm	0.02 	$&	$	19.73 	_{-	3.98 	}^{+	5.26 	}$	&	\nodata							&	$	162.99 	/	202 	$	&	\nodata							&	\nodata							&	\nodata					\\
124		& Apr 28 00:50:41.835	&	0.405	&$	0.63	\pm	0.09	$&	$	33.78	_{-	7.48	}^{+	9.89	}$	&	\nodata							&	$	212.63	/	202 	$	&	\nodata							&	\nodata							&	\nodata					\\
125		& Apr 28 00:51:35.912	&	0.069	&$	0.77 	\pm	0.04 	$&	$	16.16 	_{-	1.43 	}^{+	1.60 	}$	&	\nodata							&	$	197.26 	/	202 	$	&	\nodata							&	\nodata							&	\nodata					\\
126		& Apr 28 00:51:55.444	&	0.132	&$	0.86 	\pm	0.05 	$&	$	13.62 	_{-	1.20 	}^{+	1.34 	}$	&	\nodata							&	$	204.48 	/	202 	$	&	\nodata							&	\nodata							&	\nodata					\\
127		& Apr 28 00:52:06.141	&	0.380	&$	0.75 	\pm	0.08 	$&	$	28.08 	_{-	4.30 	}^{+	5.15 	}$	&	\nodata							&	$	194.49 	/	202 	$	&	\nodata							&	\nodata							&	\nodata					\\
128		& Apr 28 00:54:57.448	&	0.172	&$	4.45 	\pm	0.09 	$&	$	24.06 	_{-	0.71 	}^{+	0.70 	}$	&	$	-0.24 	_{-	0.17 	}^{+	0.18 	}$	&	$	344.14 	/	268 	$	&	$	4.69 	_{-	0.58 	}^{+	0.32 	}$	&	$	10.09 	_{-	1.14 	}^{+	0.97 	}$	&	$	341.98 	/	267 	$	\\
129		& Apr 28 00:56:49.646	&	0.328	&$	1.44 	\pm	0.09 	$&	\nodata							&	\nodata							&	\nodata					&	$	3.68 	_{-	0.59 	}^{+	0.56 	}$	&	$	9.46 	_{-	1.46 	}^{+	2.12 	}$	&	$	243.73 	/	198 	$	\\
130	$^{T}$	& Apr 28 01:04:03.146	&	0.062	&$	0.77 	\pm	0.05 	$&	\nodata							&	\nodata							&	\nodata					&	$	4.15 	_{-	0.56 	}^{+	0.50 	}$	&	$	11.40 	_{-	2.06 	}^{+	2.72 	}$	&	$	225.28 	/	197 	$	\\
131	$^{T}$	& Apr 28 02:00:11.518	&	0.234	&$	1.70 	\pm	0.08 	$&	\nodata							&	\nodata							&	\nodata					&	$	2.91 	_{-	0.42 	}^{+	0.42 	}$	&	$	9.57 	_{-	0.81 	}^{+	0.96 	}$	&	$	312.10 	/	267 	$	\\
132		& Apr 28 02:27:24.905	&	0.026	&$	0.20 	\pm	0.03 	$&	$	31.56 	_{-	6.35 	}^{+	8.14 	}$	&	\nodata							&	$	182.16 	/	202 	$	&	\nodata							&	\nodata							&	\nodata					\\
133		& Apr 28 03:32:00.607	&	0.130	&$	0.61 	\pm	0.07 	$&	\nodata							&	\nodata							&	\nodata					&	$	2.72 	_{-	0.51 	}^{+	0.59 	}$	&	$	14.99 	_{-	1.99 	}^{+	2.30 	}$	&	$	231.65 	/	199 	$	\\
134	$^{T}$	& Apr 28 03:47:52.140	&	0.143	&$	2.00 	\pm	0.07 	$&	$	20.79 	_{-	1.13 	}^{+	1.16 	}$	&	$	-0.42 	_{-	0.31 	}^{+	0.33 	}$	&	$	300.64 	/	270 	$	&	\nodata							&	\nodata							&	\nodata					\\
135	$^{T}$	& Apr 28 04:09:47.317	&	0.110	&$	1.89 	\pm	0.06 	$&	$	27.48 	_{-	1.61 	}^{+	1.47 	}$	&	$	-0.68 	_{-	0.23 	}^{+	0.24 	}$	&	$	358.72 	/	333 	$	&	$	4.63 	_{-	0.44 	}^{+	0.44 	}$	&	$	11.69 	_{-	0.97 	}^{+	1.18 	}$	&	$	358.28 	/	332 	$	\\
136	$^{T}$	& Apr 28 05:56:30.570	&	0.249	&$	2.21 	\pm	0.12 	$&	$	27.30 	_{-	1.43 	}^{+	1.43 	}$	&	$	0.21 	_{-	0.43 	}^{+	0.47 	}$	&	$	140.81 	/	131 	$	&	$	5.25 	_{-	2.60 	}^{+	1.26 	}$	&	$	10.09 	_{-	2.41 	}^{+	7.02 	}$	&	$	141.30 	/	130 	$	\\
137	$^{T}$	& Apr 28 09:51:04.838	&	0.240	&$	2.35 	\pm	0.10 	$&	\nodata							&	\nodata							&	\nodata					&	$	4.12 	_{-	0.30 	}^{+	0.28 	}$	&	$	11.49 	_{-	1.99 	}^{+	3.19 	}$	&	$	208.03 	/	199 	$	\\
138		& Apr 29 11:13:57.687	&	0.485	&$	0.88 	\pm	0.12 	$&	\nodata							&	$	-1.51 	_{-	0.15 	}^{+	0.14 	}$	&	$	252.61 	/	201 	$	&	\nodata							&	\nodata							&	\nodata					\\
139	$^{T}$	& Apr 29 20:47:27.860	&	0.282	&$	41.61	\pm	0.34	$&	$	34.35	_{-	0.215	}^{+	0.216	}$	&	$	0.834	_{-	0.0621	}^{+	0.0628	}$	&	$	309.9	/	200 	$	&	$	6.006	_{-	0.589	}^{+	0.531	}$	&	$	10.31	_{-	0.324	}^{+	0.405	}$	&	$	313.89	/	199 	$	\\
140	$^{T}$	& May 03 23:25:13.437	&	0.186	&$	13.69 	\pm	0.16 	$&	$	30.24 	_{-	0.36 	}^{+	0.36 	}$	&	$	0.11 	_{-	0.09 	}^{+	0.09 	}$	&	$	321.96 	/	270 	$	&	$	5.69 	_{-	0.28 	}^{+	0.26 	}$	&	$	11.45 	_{-	0.52 	}^{+	0.58 	}$	&	$	320.24 	/	269 	$	\\
141		& May 05 02:54:05.299	&	0.025	&$	0.17 	\pm	0.02 	$&	$	21.99 	_{-	4.68 	}^{+	5.84 	}$	&	\nodata							&	$	253.91 	/	270 	$	&	\nodata							&	\nodata							&	\nodata					\\
142		& May 05 03:02:56.033	&	0.163	&$	0.42 	\pm	0.06 	$&	$	51.90 	_{-	9.92 	}^{+	13.30 	}$	&	\nodata							&	$	311.04 	/	270 	$	&	\nodata							&	\nodata							&	\nodata					\\
143		& May 09 00:39:12.747	&	0.013	&$	0.30 	\pm	0.04 	$&	$	25.10 	_{-	4.36 	}^{+	5.42 	}$	&	\nodata							&	$	111.97 	/	132 	$	&	\nodata							&	\nodata							&	\nodata					\\
144	$^{T}$	& May 10 21:51:16.278	&	0.396	&$	46.49 	\pm	0.67 	$&	$	37.84 	_{-	0.49 	}^{+	0.50 	}$	&	$	0.26 	_{-	0.10 	}^{+	0.10 	}$	&	$	106.37 	/	64 	$	&	$	7.16 	_{-	0.45 	}^{+	0.43 	}$	&	$	14.22 	_{-	0.87 	}^{+	1.11 	}$	&	$	103.61 	/	63 	$	\\
145	$^{T}$	& May 19 18:32:30.295	&	0.688	&$	4.66	\pm	0.23	$&	\nodata							&	\nodata							&	\nodata					&	$	4.132	_{-	0.384	}^{+	0.396	}$	&	$	13.37	_{-	1.22	}^{+	1.41	}$	&	$	70.726	/	63 	$	\\
146		& May 19 18:57:36.305	&	0.033	&$	0.17 	\pm	0.03 	$&	\nodata							&	\nodata							&	\nodata					&	$	8.81 	_{-	1.20 	}^{+	1.36 	}$	&	\nodata							&	$	102.60 	/	132 	$	\\
147	$^{T}$	& May 20 14:10:49.826	&	0.085	&$	0.47 	\pm	0.05 	$&	$	33.64 	_{-	4.79 	}^{+	5.76 	}$	&	\nodata							&	$	267.70 	/	202 	$	&	\nodata							&	\nodata							&	\nodata					\\
148	$^{T}$	& May 20 21:47:07.495	&	0.446	&$	5.02 	\pm	0.13 	$&	$	24.37 	_{-	2.39 	}^{+	2.03 	}$	&	$	-1.37 	_{-	0.13 	}^{+	0.14 	}$	&	$	254.28 	/	202 	$	&	$	4.75 	_{-	0.22 	}^{+	0.22 	}$	&	$	15.47 	_{-	0.95 	}^{+	1.05 	}$	&	$	249.59 	/	201 	$	
\enddata
\tablecomments{$^T$ Bursts triggered GBM. \\
$^a$ Fluence in $8-200$~keV. \\
$^b$ C-Stat for the COMPT model fit or OTTB/PL fit. \\
$^c$ C-Stat for the BB+BB model fit or BB fit.}
\end{deluxetable*}
\end{longrotatetable}

\acknowledgments

We thank the anonymous reviewer for helpful comments. L.~L. acknowledges support from the National Natural Science Foundation of China (grant no. 11703002).  M.~G.~B. thanks the National Science Foundation for support under grant AST-1813649. C.\,K. acknowledges support from NASA under Fermi Guest Observer cycle-10 grant 80NSSC17K0761. Y.~K. acknowledges the support from the Scientific and Technological Research Council of Turkey (T\"{U}B\.{I}TAK grant no. 118F344).

\bibliography{refs} 

\begin{thebibliography}{}
\expandafter\ifx\csname natexlab\endcsname\relax\def\natexlab#1{#1}\fi
\providecommand{\url}[1]{\href{#1}{#1}}
\providecommand{\dodoi}[1]{doi:~\href{http://doi.org/#1}{\nolinkurl{#1}}}
\providecommand{\doeprint}[1]{\href{http://ascl.net/#1}{\nolinkurl{http://ascl.net/#1}}}
\providecommand{\doarXiv}[1]{\href{https://arxiv.org/abs/#1}{\nolinkurl{https://arxiv.org/abs/#1}}}

\bibitem[{{Ambrosi} {et~al.}(2020){Ambrosi}, {Barthelmy}, {D'Elia}, {Gropp},
  {Kennea}, {Klingler}, {Lien}, {Page}, \& {Neil Gehrels Swift Observatory
  Team}}]{gcn27672}
{Ambrosi}, E., {Barthelmy}, S.~D., {D'Elia}, V., {et~al.} 2020, GRB Coordinates
  Network, 27672, 1

\bibitem[{{Barthelmy} {et~al.}(2020){Barthelmy}, {Gropp}, {Kennea}, {Kuin},
  {Laha}, {Marshall}, {Moss}, {Page}, {Palmer}, {Sbarufatti}, {Siegel},
  {Tohuvavohu}, \& {Neil Gehrels Swift Observatory Team}}]{gcn27696}
{Barthelmy}, S.~D., {Gropp}, J.~D., {Kennea}, J.~A., {et~al.} 2020, GRB
  Coordinates Network, 27696, 1

\bibitem[{{Bochenek} {et~al.}(2020){Bochenek}, {Ravi}, {Belov}, {Hallinan},
  {Kocz}, {Kulkarni}, \& {McKenna}}]{bochenek20}
{Bochenek}, C.~D., {Ravi}, V., {Belov}, K.~V., {et~al.} 2020, arXiv e-prints,
  arXiv:2005.10828.
\newblock \doarXiv{2005.10828}

\bibitem[{{Cash}(1979)}]{cash79}
{Cash}, W. 1979, \apj, 228, 939, \dodoi{10.1086/156922}

\bibitem[{{Cherry} {et~al.}(2020){Cherry}, {Yoshida}, {Sakamoto}, {Pal'Shin},
  {Sugita}, {Kawakubo}, {Yamaoka}, {Nakahira}, {Asaoka}, {Torii}, {Shimizu},
  {Tamura}, {Cannady}, {Ricciarini}, {Marrocchesi}, \& {Calet
  Collaboration}}]{GCN27623}
{Cherry}, M.~L., {Yoshida}, A., {Sakamoto}, T., {et~al.} 2020, GRB Coordinates
  Network, 27623, 1

\bibitem[{Collazzi {et~al.}(2015)Collazzi, Kouveliotou, van~der Horst, Younes,
  Kaneko, Gö{\u{g}}ü{\c{s}}, Lin, Granot, Finger, Chaplin, Huppenkothen,
  Watts, von Kienlin, Baring, Gruber, Bhat, Gibby, Gehrels, McEnery, van~der
  Klis, \& Wijers}]{Collazzi2015}
Collazzi, A.~C., Kouveliotou, C., van~der Horst, A.~J., {et~al.} 2015, The
  Astrophysical Journal Supplement Series, 218, 11,
  \dodoi{10.1088/0067-0049/218/1/11}

\bibitem[{{Duncan} \& {Thompson}(1992)}]{dt92}
{Duncan}, R.~C., \& {Thompson}, C. 1992, \apjl, 392, L9, \dodoi{10.1086/186413}

\bibitem[{{G{\"o}{\v{g}}{\"u}{\textcommabelow s}}(2014)}]{gogus14}
{G{\"o}{\v{g}}{\"u}{\textcommabelow s}}, E. 2014, Astronomische Nachrichten,
  335, 296, \dodoi{10.1002/asna.201312035}

\bibitem[{{G{\"o}{\v{g}}{\"u}{\textcommabelow s}}
  {et~al.}(2001){G{\"o}{\v{g}}{\"u}{\textcommabelow s}}, {Kouveliotou},
  {Woods}, {Thompson}, {Duncan}, \& {Briggs}}]{gogus01}
{G{\"o}{\v{g}}{\"u}{\textcommabelow s}}, E., {Kouveliotou}, C., {Woods}, P.~M.,
  {et~al.} 2001, \apj, 558, 228, \dodoi{10.1086/322463}

\bibitem[{{Gronwall} {et~al.}(2020){Gronwall}, {Gropp}, {Kennea}, {Laha},
  {Lien}, {Marshall}, {Page}, {Palmer}, \& {Neil Gehrels Swift Observatory
  Team}}]{gcn27746}
{Gronwall}, C., {Gropp}, J.~D., {Kennea}, J.~A., {et~al.} 2020, GRB Coordinates
  Network, 27746, 1

\bibitem[{{Hurley} {et~al.}(1999){Hurley}, {Cline}, {Mazets}, {Barthelmy},
  {Butterworth}, {Marshall}, {Palmer}, {Aptekar}, {Golenetskii}, {Il'Inskii},
  {Frederiks}, {McTiernan}, {Gold}, \& {Trombka}}]{hurley99}
{Hurley}, K., {Cline}, T., {Mazets}, E., {et~al.} 1999, \nat, 397, 41,
  \dodoi{10.1038/16199}

\bibitem[{{Hurley} {et~al.}(2020){Hurley}, {Mitrofanov}, {Golovin}, {Litvak},
  {Sanin}, {Kozlova}, {Golenetskii}, {Aptekar}, {Frederiks}, {Svinkin},
  {Cline}, {Goldstein}, {Briggs}, {Wilson-Hodge}, {von Kienlin}, {Zhang},
  {Rau}, {Savchenko}, {E. Bozzo}, {Ferrigno}, {Barthelmy}, {Cummings}, {Krimm},
  {Palmer}, {Boynton}, {Fellows}, {Harshman}, {Enos}, \& {Starr}}]{GCN27625}
{Hurley}, K., {Mitrofanov}, I.~G., {Golovin}, D., {et~al.} 2020, GRB
  Coordinates Network, 27625, 1

\bibitem[{{Israel} {et~al.}(2008){Israel}, {Romano}, {Mangano}, {Dall'Osso},
  {Chincarini}, {Stella}, {Campana}, {Belloni}, {Tagliaferri}, {Blustin},
  {Sakamoto}, {Hurley}, {Zane}, {Moretti}, {Palmer}, {Guidorzi}, {Burrows},
  {Gehrels}, \& {Krimm}}]{israel2008}
{Israel}, G.~L., {Romano}, P., {Mangano}, V., {et~al.} 2008, \apj, 685, 1114,
  \dodoi{10.1086/590486}

\bibitem[{{Israel} {et~al.}(2016){Israel}, {Esposito}, {Rea}, {Coti Zelati},
  {Tiengo}, {Campana}, {Mereghetti}, {Rodriguez Castillo}, {G{\"o}tz},
  {Burgay}, {Possenti}, {Zane}, {Turolla}, {Perna}, {Cannizzaro}, \&
  {Pons}}]{israel16}
{Israel}, G.~L., {Esposito}, P., {Rea}, N., {et~al.} 2016, \mnras, 457, 3448,
  \dodoi{10.1093/mnras/stw008}

\bibitem[{{Kaspi} \& {Beloborodov}(2017)}]{kas17}
{Kaspi}, V.~M., \& {Beloborodov}, A.~M. 2017, \araa, 55, 261,
  \dodoi{10.1146/annurev-astro-081915-023329}

\bibitem[{{Kirsten} {et~al.}(2020){Kirsten}, {Snelders}, {Jenkins}, {Nimmo},
  {den van Eijnden}, {Hessels}, {Gawronski}, \& {Yang}}]{2020arXiv200705101K}
{Kirsten}, F., {Snelders}, M., {Jenkins}, M., {et~al.} 2020, arXiv e-prints,
  arXiv:2007.05101.
\newblock \doarXiv{2007.05101}

\bibitem[{{Kouveliotou} {et~al.}(1993){Kouveliotou}, {Meegan}, {Fishman},
  {Bhat}, {Briggs}, {Koshut}, {Paciesas}, \& {Pendleton}}]{Kouveliotou1993}
{Kouveliotou}, C., {Meegan}, C.~A., {Fishman}, G.~J., {et~al.} 1993, \apjl,
  413, L101, \dodoi{10.1086/186969}

\bibitem[{{Kouveliotou} {et~al.}(1998){Kouveliotou}, {Dieters}, {Strohmayer},
  {van Paradijs}, {Fishman}, {Meegan}, {Hurley}, {Kommers}, {Smith}, {Frail},
  \& {Murakami}}]{ck1998}
{Kouveliotou}, C., {Dieters}, S., {Strohmayer}, T., {et~al.} 1998, \nat, 393,
  235, \dodoi{10.1038/30410}

\bibitem[{{Li} {et~al.}(2020{\natexlab{a}}){Li}, {Lin}, {Xiong}, {Ge}, {Li},
  {Li}, {Lu}, {Zhang}, {Tuo}, {Nang}, {Zhang}, {Xiao}, {Chen}, {Song}, {Xu},
  {Liu}, {Jia}, {Cao}, {Zhang}, {Qu}, {Liao}, {Zhao}, {Tan}, {Nie}, {Zhao},
  {Zheng}, {Zheng}, {Luo}, {Cai}, {Li}, {Xue}, {Bu}, {Chang}, {Chen}, {Chen},
  {Chen}, {Chen}, {Chen}, {Cui}, {Cui}, {Deng}, {Dong}, {Du}, {Fu}, {Gao},
  {Gao}, {Gao}, {Gu}, {Guan}, {Guo}, {Han}, {Huang}, {Huo}, {Jiang}, {Jiang},
  {Jin}, {Jin}, {Kong}, {Li}, {Li}, {Li}, {Li}, {Li}, {Li}, {Li}, {Liang},
  {Liu}, {Liu}, {Liu}, {Liu}, {Liu}, {Lu}, {Lu}, {Luo}, {Ma}, {Meng}, {Ou},
  {Sai}, {Shang}, {Song}, {Sun}, {Tao}, {Wang}, {Wang}, {Wang}, {Wang}, {Wang},
  {Wen}, {Wu}, {Wu}, {Wu}, {Xiao}, {Yang}, {Yang}, {Yang}, {Yang}, {Yi}, {Yin},
  {You}, {Zhang}, {Zhang}, {Zhang}, {Zhang}, {Zhang}, {Zhang}, {Zhang},
  {Zhang}, {Zhang}, {Zhang}, {Zhang}, {Zhang}, {Zhang}, {Zhang}, {Zhang},
  {Zhang}, {Zhou}, {Zhou}, {Zhu}, {Zhu}, \& {Zhuang}}]{li2020}
{Li}, C.~K., {Lin}, L., {Xiong}, S.~L., {et~al.} 2020{\natexlab{a}}, arXiv
  e-prints, arXiv:2005.11071.
\newblock \doarXiv{2005.11071}

\bibitem[{{Li} {et~al.}(2020{\natexlab{b}}){Li}, {Tuo}, {Ge}, {Li}, {Liao},
  {Cai}, {Xiong}, {Liu}, {Chen}, {Cao}, {Li}, {Jia}, {Nie}, {Lu}, {Song}, {Wu},
  {Xu}, {Zhang}, {Lin}, {Zhang}, \& {Insight-HXMT Team}}]{2020GCN.27679....1L}
{Li}, C.~K., {Tuo}, Y.~L., {Ge}, M.~Y., {et~al.} 2020{\natexlab{b}}, GRB
  Coordinates Network, 27679, 1

\bibitem[{Lin {et~al.}(2020)Lin, Gö{\u{g}}ü{\c{s}}, Roberts, Kouveliotou,
  Kaneko, van~der Horst, \& Younes}]{Lin2020}
Lin, L., Gö{\u{g}}ü{\c{s}}, E., Roberts, O.~J., {et~al.} 2020, The
  Astrophysical Journal, 893, 156, \dodoi{10.3847/1538-4357/ab818f}

\bibitem[{{Lin} {et~al.}(2011){Lin}, {Kouveliotou},
  {G{\"o}{\v{g}}{\"u}{\textcommabelow s}}, {van der Horst}, {Watts}, {Baring},
  {Kaneko}, {Wijers}, {Woods}, {Barthelmy}, {Burgess}, {Chaplin}, {Gehrels},
  {Goldstein}, {Granot}, {Guiriec}, {Mcenery}, {Preece}, {Tierney}, {van der
  Klis}, {von Kienlin}, \& {Zhang}}]{lin11}
{Lin}, L., {Kouveliotou}, C., {G{\"o}{\v{g}}{\"u}{\textcommabelow s}}, E.,
  {et~al.} 2011, \apjl, 740, L16, \dodoi{10.1088/2041-8205/740/1/L16}

\bibitem[{{Lin} {et~al.}(2012){Lin}, {G{\"o}{\v{g}}{\"u}{\textcommabelow s}},
  {Baring}, {Granot}, {Kouveliotou}, {Kaneko}, {van der Horst}, {Gruber}, {von
  Kienlin}, {Younes}, {Watts}, \& {Gehrels}}]{lin2012}
{Lin}, L., {G{\"o}{\v{g}}{\"u}{\textcommabelow s}}, E., {Baring}, M.~G.,
  {et~al.} 2012, \apj, 756, 54, \dodoi{10.1088/0004-637X/756/1/54}

\bibitem[{{Lin} {et~al.}(2020){Lin}, {Zhang}, {Wang}, {Gao}, {Guan}, {Han},
  {Jiang}, {Jiang}, {Lee}, {Li}, {Men}, {Miao}, {Niu}, {Niu}, {Sun}, {Wang},
  {Wang}, {Xu}, {Xu}, {Xu}, {Yang}, {Yang}, {Yu}, {Zhang}, {Zhang}, {Zhou},
  {Zhu}, {Castro-Tirado}, {Dai}, {Ge}, {Hu}, {Li}, {Li}, {Li}, {Liang}, {Jia},
  {Querel}, {Shao}, {Wang}, {Wang}, {Wu}, {Xiong}, {Xu}, {Yang}, {Zhang},
  {Zhang}, {Zheng}, \& {Zou}}]{Lin2020b}
{Lin}, L., {Zhang}, C.~F., {Wang}, P., {et~al.} 2020, arXiv e-prints,
  arXiv:2005.11479.
\newblock \doarXiv{2005.11479}

\bibitem[{{Meegan} {et~al.}(2009){Meegan}, {Lichti}, {Bhat}, {Bissaldi},
  {Briggs}, {Connaughton}, {Diehl}, {Fishman}, {Greiner}, {Hoover}, {van der
  Horst}, {von Kienlin}, {Kippen}, {Kouveliotou}, {McBreen}, {Paciesas},
  {Preece}, {Steinle}, {Wallace}, {Wilson}, \& {Wilson-Hodge}}]{meegan2009}
{Meegan}, C., {Lichti}, G., {Bhat}, P.~N., {et~al.} 2009, \apj, 702, 791,
  \dodoi{10.1088/0004-637X/702/1/791}

\bibitem[{{Mereghetti} {et~al.}(2020){Mereghetti}, {Savchenko}, {Ferrigno},
  {G{\"o}tz}, {Rigoselli}, {Tiengo}, {Bazzano}, {Bozzo}, {Coleiro},
  {Courvoisier}, {Doyle}, {Goldwurm}, {Hanlon}, {Jourdain}, {von Kienlin},
  {Lutovinov}, {Martin-Carrillo}, {Molkov}, {Natalucci}, {Onori}, {Panessa},
  {Rodi}, {Rodriguez}, {S{\'a}nchez-Fern{\'a}ndez}, {Sunyaev}, \&
  {Ubertini}}]{2020ApJ...898L..29M}
{Mereghetti}, S., {Savchenko}, V., {Ferrigno}, C., {et~al.} 2020, \apjl, 898,
  L29, \dodoi{10.3847/2041-8213/aba2cf}

\bibitem[{{Olausen} \& {Kaspi}(2014)}]{olausen14}
{Olausen}, S.~A., \& {Kaspi}, V.~M. 2014, \apjs, 212, 6,
  \dodoi{10.1088/0067-0049/212/1/6}

\bibitem[{{Palmer}(2020)}]{palmer20}
{Palmer}, D.~M. 2020, The Astronomer's Telegram, 13675, 1

\bibitem[{{Palmer} {et~al.}(2005){Palmer}, {Barthelmy}, {Gehrels}, {Kippen},
  {Cayton}, {Kouveliotou}, {Eichler}, {Wijers}, {Woods}, {Granot}, {Lyubarsky},
  {Ramirez-Ruiz}, {Barbier}, {Chester}, {Cummings}, {Fenimore}, {Finger},
  {Gaensler}, {Hullinger}, {Krimm}, {Markwardt}, {Nousek}, {Parsons}, {Patel},
  {Sakamoto}, {Sato}, {Suzuki}, \& {Tueller}}]{palmer2005}
{Palmer}, D.~M., {Barthelmy}, S., {Gehrels}, N., {et~al.} 2005, \nat, 434,
  1107, \dodoi{10.1038/nature03525}

\bibitem[{{Ridnaia} {et~al.}(2020{\natexlab{a}}){Ridnaia}, {Golenetskii},
  {Aptekar}, {Frederiks}, {Ulanov}, {Svinkin}, {Tsvetkova}, {Lysenko}, {Cline},
  \& {Konus-Wind Team}}]{GCN27631}
{Ridnaia}, A., {Golenetskii}, S., {Aptekar}, R., {et~al.} 2020{\natexlab{a}},
  GRB Coordinates Network, 27631, 1

\bibitem[{{Ridnaia} {et~al.}(2020{\natexlab{b}}){Ridnaia}, {Svinkin},
  {Frederiks}, {Bykov}, {Popov}, {Aptekar}, {Golenetskii}, {Lysenko},
  {Tsvetkova}, {Ulanov}, \& {Cline}}]{kw200428}
{Ridnaia}, A., {Svinkin}, D., {Frederiks}, D., {et~al.} 2020{\natexlab{b}},
  arXiv e-prints, arXiv:2005.11178.
\newblock \doarXiv{2005.11178}

\bibitem[{Scargle {et~al.}(2013)Scargle, Norris, Jackson, \&
  Chiang}]{Scargle2013}
Scargle, J.~D., Norris, J.~P., Jackson, B., \& Chiang, J. 2013, The
  Astrophysical Journal, 764, 167, \dodoi{10.1088/0004-637x/764/2/167}

\bibitem[{{Stamatikos} {et~al.}(2014){Stamatikos}, {Malesani}, {Page}, \&
  {Sakamoto}}]{mstam14}
{Stamatikos}, M., {Malesani}, D., {Page}, K.~L., \& {Sakamoto}, T. 2014, GRB
  Coordinates Network, 16520, 1

\bibitem[{{The CHIME/FRB Collaboration} {et~al.}(2020){The CHIME/FRB
  Collaboration}, {:}, {Andersen}, {Band ura}, {Bhardwaj}, {Bij}, {Boyce},
  {Boyle}, {Brar}, {Cassanelli}, {Chawla}, {Chen}, {Cliche}, {Cook},
  {Cubranic}, {Curtin}, {Denman}, {Dobbs}, {Dong}, {Fandino}, {Fonseca},
  {Gaensler}, {Giri}, {Good}, {Halpern}, {Hill}, {Hinshaw}, {H{\"o}fer},
  {Josephy}, {Kania}, {Kaspi}, {Landecker}, {Leung}, {Li}, {Lin}, {Masui},
  {Mckinven}, {Mena-Parra}, {Merryfield}, {Meyers}, {Michilli}, {Milutinovic},
  {Mirhosseini}, {M{\"u}nchmeyer}, {Naidu}, {Newburgh}, {Ng}, {Patel}, {Pen},
  {Pinsonneault-Marotte}, {Pleunis}, {Quine}, {Rafiei-Ravandi}, {Rahman},
  {Ransom}, {Renard}, {Sanghavi}, {Scholz}, {Shaw}, {Shin}, {Siegel}, {Singh},
  {Smegal}, {Smith}, {Stairs}, {Tan}, {Tendulkar}, {Tretyakov}, {Vanderlinde},
  {Wang}, {Wulf}, \& {Zwaniga}}]{scholz20}
{The CHIME/FRB Collaboration}, {:}, {Andersen}, B.~C., {et~al.} 2020, arXiv
  e-prints, arXiv:2005.10324.
\newblock \doarXiv{2005.10324}

\bibitem[{{Thompson} \& {Duncan}(1995)}]{td95}
{Thompson}, C., \& {Duncan}, R.~C. 1995, \mnras, 275, 255,
  \dodoi{10.1093/mnras/275.2.255}

\bibitem[{{van der Horst} {et~al.}(2012){van der Horst}, {Kouveliotou},
  {Gorgone}, {Kaneko}, {Baring}, {Guiriec}, {G{\"o}{\v{g}}{\"u}{\textcommabelow
  s}}, {Granot}, {Watts}, {Lin}, {Bhat}, {Bissaldi}, {Chaplin}, {Finger},
  {Gehrels}, {Gibby}, {Giles}, {Goldstein}, {Gruber}, {Harding}, {Kaper}, {von
  Kienlin}, {van der Klis}, {McBreen}, {Mcenery}, {Meegan}, {Paciesas},
  {Pe'er}, {Preece}, {Ramirez-Ruiz}, {Rau}, {Wachter}, {Wilson-Hodge}, {Woods},
  \& {Wijers}}]{AvdH2012}
{van der Horst}, A.~J., {Kouveliotou}, C., {Gorgone}, N.~M., {et~al.} 2012,
  \apj, 749, 122, \dodoi{10.1088/0004-637X/749/2/122}

\bibitem[{von Kienlin {et~al.}(2012)von Kienlin, Gruber, Kouveliotou, Granot,
  Baring, Gö{\u{g}}ü{\c{s}}, Huppenkothen, Kaneko, Lin, Watts, Bhat, Guiriec,
  van~der Horst, Bissaldi, Greiner, Meegan, Paciesas, Preece, \&
  Rau}]{vonKienlin2012}
von Kienlin, A., Gruber, D., Kouveliotou, C., {et~al.} 2012, The Astrophysical
  Journal, 755, 150, \dodoi{10.1088/0004-637x/755/2/150}

\bibitem[{{Younes} {et~al.}(2017){Younes}, {Kouveliotou}, {Jaodand}, {Baring},
  {van der Horst}, {Harding}, {Hessels}, {Gehrels}, {Gill}, {Huppenkothen},
  {Granot}, {G{\"o}{\u{g}}{\"u}{\textcommabelow s}}, \& {Lin}}]{younes17}
{Younes}, G., {Kouveliotou}, C., {Jaodand}, A., {et~al.} 2017, \apj, 847, 85,
  \dodoi{10.3847/1538-4357/aa899a}

\bibitem[{{Younes} {et~al.}(2020{\natexlab{a}}){Younes}, {Guver}, {Enoto},
  {Arzoumanian}, {Gendreau}, {Hu}, {Ray}, {Kouveliotou}, {Guillot}, {Ho},
  {Ferrara}, \& {Malacaria}}]{younes20}
{Younes}, G., {Guver}, T., {Enoto}, T., {et~al.} 2020{\natexlab{a}}, The
  Astronomer's Telegram, 13678, 1

\bibitem[{{Younes} {et~al.}(2020{\natexlab{b}}){Younes}, {Baring},
  {Kouveliotou}, {Arzoumanian}, {Enoto}, {Doty}, {Gendreau},
  {G{\"o}{\u{g}}{\"u}{\c{s}}}, {Guillot}, {G{\"u}ver}, {Harding}, {Ho}, {van
  der Horst}, {Jaisawal}, {Kaneko}, {LaMarr}, {Lin}, {Majid}, {Okajima},
  {Pope}, {Ray}, {Roberts}, {Saylor}, {Steiner}, \& {Wadiasingh}}]{younes20_2}
{Younes}, G., {Baring}, M.~G., {Kouveliotou}, C., {et~al.} 2020{\natexlab{b}},
  arXiv e-prints, arXiv:2006.11358.
\newblock \doarXiv{2006.11358}

\bibitem[{{Zhang} {et~al.}(2020){Zhang}, {Jiang}, {Men}, {Wang}, {Xu}, {Xu},
  {Niu}, {Zhou}, {Guan}, {Han}, {Jiang}, {Lee}, {Li}, {Lin}, {Niu}, {Wang},
  {Wang}, {Xu}, {Yu}, {Zhang}, \& {Zhu}}]{fastatel2020b}
{Zhang}, C.~F., {Jiang}, J.~C., {Men}, Y.~P., {et~al.} 2020, The Astronomer's
  Telegram, 13699, 1

\end{thebibliography}

\end{document}